\documentclass{aa}
\usepackage{graphicx}
\usepackage{natbib}
\usepackage{subfig}

\usepackage[dvips,dvipsnames,usenames]{color}
\usepackage{varioref}
\usepackage{colortbl}
\usepackage[dvips, breaklinks={true}, bookmarks, colorlinks={true}, pdfpagemode={NON}, pdffitwindow=true, pdfstartview={FitB}, linkcolor={BlueViolet}, menucolor={BlueViolet}, citecolor={Mahogany}, filecolor={OliveGreen}, pagecolor={Mahogany}, urlcolor={MidnightBlue}, pdftitle={CLASH-photoz}, pdfauthor={Stephanie Jouvel}, pdfsubject={Calibration}, pdfkeywords={CLASH-photoz}]{hyperref}
\usepackage[all]{hypcap}
\bibpunct{(}{)}{;}{a}{}{,}

\begin{document}
\title{CLASH: Photometric redshifts with 16 HST bands in galaxy cluster fields\thanks{Online-only data are at the CDS via anonymous ftp to cdsarc.u-strasbg.fr (130.79.128.5) or via http://cdsweb.u-strasbg.fr/cgi-bin/qcat?J/A+A/}} 
\author{S. Jouvel \inst{1,2}, O. Host \inst{3}, O. Lahav \inst{2}, S. Seitz \inst{6}, A. Molino \inst{9},
D. Coe \inst{4}, M. Postman \inst{4}, L. Moustakas \inst{19}, N. Ben\`itez \inst{9}, 
P. Rosati \inst{5}, I. Balestra \inst{21,25},
C. Grillo \inst{3}, L. Bradley \inst{4}, A. Fritz \inst{23}, D. Kelson \inst{16}, A. M. Koekemoer \inst{4}, 
D. Lemze \inst{10}, E. Medezinski \inst{10}, A. Mercurio \inst{21},  J. Moustakas \inst{22}, M. Nonino \inst{25},  
M. Scodeggio \inst{23}, W. Zheng \inst{10}, A. Zitrin \inst{24,26}, M. Bartelmann \inst{11}, R. Bouwens \inst{12}, 
T. Broadhurst \inst{7,8}, M. Donahue \inst{13}, H. Ford \inst{10}, 
G. Graves \inst{14},  L. Infante \inst{15}, Y. Jimenez-Teja \inst{9}, 
R. Lazkoz \inst{7}, P. Melchior \inst{17}, M. Meneghetti \inst{18}, J. Merten \inst{19}, S. Ogaz \inst{4}, K. Umetsu \inst{20}} 
  \institute{
  Institut de Ci\`encies de l'Espai (IEEC-CSIC), E-08193 Bellaterra (Barcelona), Spain 
  \and University College London, Gower street, London WC1E 6BT, UK 
  \and Dark Cosmology Centre, Niels Bohr Institute, University of Copenhagen, Juliane Maries Vej 30, DK-2100 Copenhagen, Denmark 
  \and Space Telescope Science Institute, 3700 San Martin Drive, Baltimore, MD 21218, USA 
  \and European Southern Observatory (ESO), D-85748 Garching, Germany 
  \and Instituts für Astronomie und Astrophysik, Universitäs-Sternwarte München, D-81679 München, Germany
  \and Department of Theoretical Physics, University of the Basque Country UPV/EHU, E-48080 Bilbao, Spain
  \and Ikerbasque, Basque Foundation for Science, E-48011 Bilbao, Spain
  \and Instituto de Astrof´ısica de Andaluc´ıa (IAA-CSIC), E-18008 Granada, Spain
  \and Department of Physics and Astronomy, The Johns Hopkins University, Baltimore, MD, USA
  \and Institut für Theoretische Astrophysik, Zentrum für Astronomie, Institut für Theoretische Astrophysik, Albert-Ueberle-Str. 2, D-29120 Heidelberg, Germany
  \and Leiden Observatory, Leiden University, NL-2333 Leiden, The Netherlands
  \and Department of Physics and Astronomy, Michigan State University, East Lansing, MI, USA
  \and Department of Astronomy, University of California, Berkeley, CA 94720, USA
  \and Instituto de Astrofísica y AIUC, P. Universidad Católica de Chile
  \and Carnegie Observatories, Carnegie Institute for Science, Pasadena, CA, USA
  \and Center for Cosmology and Astro-Particle Physics, The Ohio State University, 191 W. Woodruff Ave., Columbus, Ohio 43210, USA
  \and INAF, Osservatorio Astronomico di Bologna; INFN, Sezione di Bologna
  \and Jet Propulsion Laboratory, California Institute of Technology
  \and Institute of Astronomy and Astrophysics, Academia Sinica, P.O. Box 23-141, Taipei 10617, Taiwan
  \and INAF - Osservatorio Astronomico di Capodimonte, via Moiariello 16, I-80131 Napoli, Italy
  \and Department of Physics and Astronomy, Siena College, 515 Loudon Road, Loudonville, NY 12211
  \and INAF - Istituto di Astrofisica Spaziale e Fisica cosmica (IASF) Milano, via Bassini 15, 20133 Milano, Italy
  \and Cahill Center for Astronomy and Astrophysics, California Institute of Technology, MS 249-17, Pasadena, CA 91125, USA 
  \and INAF - Osservatorio Astronomico di Trieste, via G.B. Tiepolo 11, 40131 Trieste, Italy
  \and Hubble Fellow 
  }
\authorrunning{St\'ephanie Jouvel et al. }
\titlerunning{Photo-z for the CLASH survey}
\offprints{St\'ephanie Jouvel, \email{jouvel@ice.cat}}
\abstract{
The Cluster Lensing And Supernovae survey with Hubble (CLASH) is Hubble Space Telescope (HST) Multi-Cycle Treasury programme 
that observes 25 massive galaxy clusters, 20 of which were X-ray-selected to preferably choose dynamically relaxed clusters
and 5 additional ``high magnification" clusters which were selected based on their optical lensing properties. 
CLASH aims to study the
dark matter distribution of the clusters and find magnified high redshift galaxies behind them.
CLASH observations were carried out in 16 bands from UV to NIR   
to derive accurate and reliable estimates of photometric redshifts. 
  }{
We present the CLASH photometric redshifts using 16 HST bands and study
the photometric redshift accuracy including a detailed comparison between photometric
and spectroscopic redshifts for the strong lensing arcs using the measurements from the
cluster MACSJ1206.2-0847.
  }{
We used the publicly available Le Phare and BPZ photometric redshift estimation codes on
17 CLASH galaxy clusters for which the full photo-z data processing had been completed at
the time of this analyses and derive an estimate of the CLASH photo-z accuracy. 
  }{
Using Le Phare code for objects with $S/N\geq10$, we reach a precision of 3\%(1+z) for the strong lensing arcs, which is reduced to 2.4\%(1+z) 
after removing outliers. For galaxies in the cluster field, the corresponding values are 4\%(1+z) 
and 3\%(1+z). Using mock galaxy catalogues, we show that 
3\%(1+z) precision is what is expected using the baseline sky substraction algorithm  
when taking into account extinction from dust, emission lines, and the finite range of SEDs 
included in  the photo-z template library. An improved method for estimating galaxy colours that 
yields more accurate photometric redshifts will be explored in a forthcoming paper.
We study photo-z results for different aperture photometry techniques
and find that the SExtractor isophotal photometry works best. We check the robustness of the arcs photo-z results by rederiving
the input photometry in the case of MACS1206. 
We describe and release a photometric redshift catalogue of the MACS1206 cluster we study. 
  }{
Our photo-z codes give similar 
results for the strong lensing arcs, as well as for galaxies of the cluster field. 
Results are improved when optimizing the photometric aperture shape that shows an optimal aperture size
around 1" radius, giving results that are equivalent to isophotal photometry. Tailored photometry of the
arcs improves the photo-z results by showing more consistency between the different arcs of the same
strong lensing system. 
}
\date{arXiv:1308.0063, accepted to A\&A the 22 nov 2013}
\maketitle
\section{Introduction}
\label{sec:intro}

CLASH \citep{Postman12} \footnote{http://www.stsci.edu/~postman/CLASH/Home.html} aims to 
provide very robust constraints on the dark matter distribution (including the central 
mass concentration parameter) in 25 massive galaxy clusters by using 16-band imaging from 
the Hubble Space Telescope (HST) and wide-field imaging from the Subaru Telescope.
These data allow joint constraints to 
be made on the mass profile and matter distribution using both strong and 
weak lensing measurements \citep{Zitrin12,Zitrin11,Umetsu12,Coe12}. A key requirement in modelling cluster gravitational 
lenses is the ability to distinguish background galaxies (those sources at 
greater redshift than the cluster that can be strongly or weakly lensed) 
from the foreground galaxies (at lower redshifts than the cluster) and from the cluster member galaxies. 
The cluster member  galaxies are important in establishing an initial guess at the cluster mass model. 
Good redshift estimates are required for both the cluster members and the 
background galaxies in order to measure the relative distance ratios between the 
lens and the sources - a key input into the lens mass modelling process. 
The 16-band HST imaging obtained in CLASH allows, in principle, superb 
photometric redshift estimates to be measured down to flux levels well 
beyond the reach of spectrographs on ground-based telescopes. 
In this paper, we quantify the accuracy of the CLASH photometric redshifts 
as well as discuss the accuracy required to reliably measure the 
cluster mass distributions. We will show that the two accuracies are in good agreement.

Redshift estimates are a necessary ingredient for gravitational lensing analyses, both weak and strong. 
Apart from identifying galaxies located behind the lensing cluster itself, the strength of the 
lensing signal depends on the ratio of the distances between lens and source galaxy and between the 
observer and the source, $D_\mathrm{LS}/D_\mathrm{S}$. In the strong lensing regime, 
accurate redshifts of each individual source are needed both to identify possible counter-images 
and to constrain the lensing mass model. For weak lensing, the mass model depends on an 
unbiased estimate of the mean distance ratio $\langle D_\mathrm{LS}/D_\mathrm{S}\rangle$ 
averaged over the background sources. Removing interlopers from the sample of background sources is paramount.

Most of the galaxies which are interesting for the lensing measurement both strong
and weak are too faint for spectroscopic observations. 
Photometric redshifts have proved to be useful 
and essential tools for a range of astronomical applications, 
from wide-field 
galaxy surveys (e.g.~the Dark Energy Survey \footnote{http://www.darkenergysurvey.org/}, 
the Large Synoptic Survey Telescope \footnote{http://www.lsst.org/lsst/}, 
and Euclid \footnote{http://sci.esa.int/euclid/}) to targeted deep field observations 
such as the COSMOS survey \citep{Scoville07,Ilbert09}. There are several different publicly available photo-z codes which, 
broadly speaking, can be categorized in two groups:
empirical training and template matching methods. The empirical training methods
include neural networks such as ANNz \citep{Collister04,Vanzella04}, decision trees 
such as ArborZ \citep{Gerdes10}, kd-trees \citep{Csabai07}. 
Training methods require a statistically significant and representative redshift-magnitude-colour 
training sample.
Template fitting methods, such as Hyper-z \citep{Bolzonella00}, BPZ \citep{Benitez00}, 
Le Phare \citep{Ilbert06}, and EAZY \citep{Brammer08}, require a representative 
library of galaxy spectral energy distributions (SEDs) that
is fitted to the observations to find the best combination of
redshift and template. 
The SED library can either be empirically 
\citep{Coleman80} or theoretically \citep{BC03} based. 

There is no empirical training set available for the CLASH data and hence we make use 
of two template-based codes, BPZ and Le Phare. \citet{Hildebrandt10} compared the photo-z 
performances on simulated and real data and both BPZ and Le Phare compared favorably. 

There have been earlier studies of galaxy photometric redshift in cluster
environment such as \citet{Pello09,Guennou10}. \citet{Guennou10} reaches a precision
of $\sigma\approx$ 0.05 (NMAD\footnote{$1.48*median[\Delta z_p /(1+z_s)]$}) 
at m(I)$<$24.5 using 5 optical ground-based broadbands BVRIZ complemented with Spitzer IRAC 3.6 and 4.5 $\mu$m
and F814W ACS/HST observations. Their study shows a degradation in the photometric redshift accuracy for
intrinsically bright galaxies and for the latest and earliest galaxy types in galaxy clusters. \citet{Guennou10} 
showed an environmental dependence of the photo-z accuracy which is interpreted as the photo-z
template being not well suited for galaxy clusters as quoted. \citet{Pello09} derived
photo-z for the ESO Distant Cluster Survey (EDisCS) fields \citep{White05} which is a sample of 20 clusters. 
They reach a precision 
$\sigma\approx$ 0.05 to 0.06 at I$<$22 with respectively 4 and 5 optical and NIR ground-based broadbands BVIKs and VRIJKs 
for respectively a low-z and high-z cluster sample. The scope of these
papers was to present their photo-z quality with applications to studies of galaxy evolution for \citet{Pello09}
and weak-lensing tomography for \citet{Guennou10}. 
In this paper, we study the accuracy of photo-z's based on HST photometry of both the cluster 
member galaxies and the lensed arcs identified as part of the CLASH strong lensing analysis. 

The paper is structured as follows: In section \ref{sec:clash}, we give a quick overview of the scientific goals and observations of the CLASH
programme and specify the role and impact of the photo-z. 
Section \ref{sec:code} presents the two photo-z techniques we use 
and we study their performance in section \ref{sec:accuracy}. 
In section \ref{subsec:simul}, we compare the performance to an idealized scenario based on mock galaxy catalogues.
Section \ref{sec:arcs} discusses issues with obtaining photo-z's for strongly lensed multiple images. 
We focus on MACS1206 as it has a high numbers of arcs and spectroscopic redshifts.
Finally we present a photo-z catalogue for MACS1206 which includes both BPZ and Le Phare photo-z.

Throughout this paper we assume a flat Lambda-CDM cosmology with $\Omega_m,\Omega_{\Lambda}=(0.3,0.7)$ 
and use the AB magnitude system.

\section{The CLASH programme}
\label{sec:clash}

CLASH is an HST Multi-Cycle Treasury programme. Observations in the 16 HST broad passbands have been completed 
for the 25 clusters. 

\subsection{The CLASH photometry}
\label{subsec:clash_photometry}

The photometry is the most critical issue for photo-z accuracy. This section gives an overview 
of the photometric quality and completeness of the CLASH data and catalogues. For more details, the reader
should refer to \citep{Postman12} section 5.

Image co-addition and alignment onto a grid with 0.065" per pixel is performed with an automated pipeline based on MosaicDrizzle 
\citep{Koekemoer11}. Object detection and photometry is then accomplished using 
SExtractor in dual image mode using 
a weighted sum of the ACS/WFC and WFC3/IR images \citep{Bertin96}. 

The photometry is measured in isophotal apertures and corrected for  
galactic extinction using \citet{Schlegel98}. 
A local sky background 
is estimated for each object within an annulus of 26 pixels. 
We do not model and subtract the galaxy cluster light from the standard pipeline products. 
The impact of a more advanced cluster light sky subtraction algorithm on the CLASH photometric 
redshift accuracy will be investigated in separate paper \citep{Molino13}.  


In Table \ref{tab:depth} we list the filters used for the CLASH 
observations with their $5\sigma$ depths derived from all the CLASH clusters sample
and the percentage of galaxies observed for each filter down to S/N(F814W)$=$5 in MACS1206. 
The magnitude limits shown are {\bf not} the point source limits but rather the 
mean limits averaged over all extended sources (i.e., galaxies)
without regards to size. Magnitude limits based on point sources can be found in \citet{Postman12}.

\begin{table}[!ht]
\caption{HST Camera, band, magnitude limits for extended sources at 5$\sigma$ and completeness for the CLASH observations.
The magnitudes limits are an average over galaxies of different sizes. The filter transmission curves are shown in \autoref{fig:clashfilters}.}
\begin{tabular}{cccccccc} \hline\hline
Camera & \# Filter & AB mag (5$\sigma$)  & \% S/N$>$5  \\
\hline\hline
WFC3/UVIS & F225W  & 24.8 &  69.3 \\
WFC3/UVIS & F275W  & 24.9 &  70.6 \\
WFC3/UVIS & F336W  & 25.2 &  79.0 \\
WFC3/UVIS & F390W  & 26.1 &  92.6 \\
ACS/WFC   & F435W  & 26.2 &  94.3 \\
ACS/WFC   & F475W  & 26.7 &  95.6 \\
ACS/WFC   & F606W  & 27.2 &  98.6 \\
ACS/WFC   & F625W  & 26.6 &  98.6 \\
ACS/WFC   & F775W  & 26.5 &  99.7 \\
ACS/WFC   & F814W  & 27.5 &  99.6 \\
ACS/WFC   & F850LP & 26.2 &  99.5 \\
WFC3/IR	  & F105W  & 27.0 &  96.2 \\	
WFC3/IR	  & F110W  & 27.6 &  84.8 \\	
WFC3/IR	  & F125W  & 27.1 &  86.7 \\	
WFC3/IR	  & F140W  & 27.3 &  95.1 \\	
WFC3/IR	  & F160W  & 27.2 &  98.0 \\	
\end{tabular}
\label{tab:depth}
\end{table}

\begin{figure*}[!t]
\centering
\includegraphics[width=\textwidth]{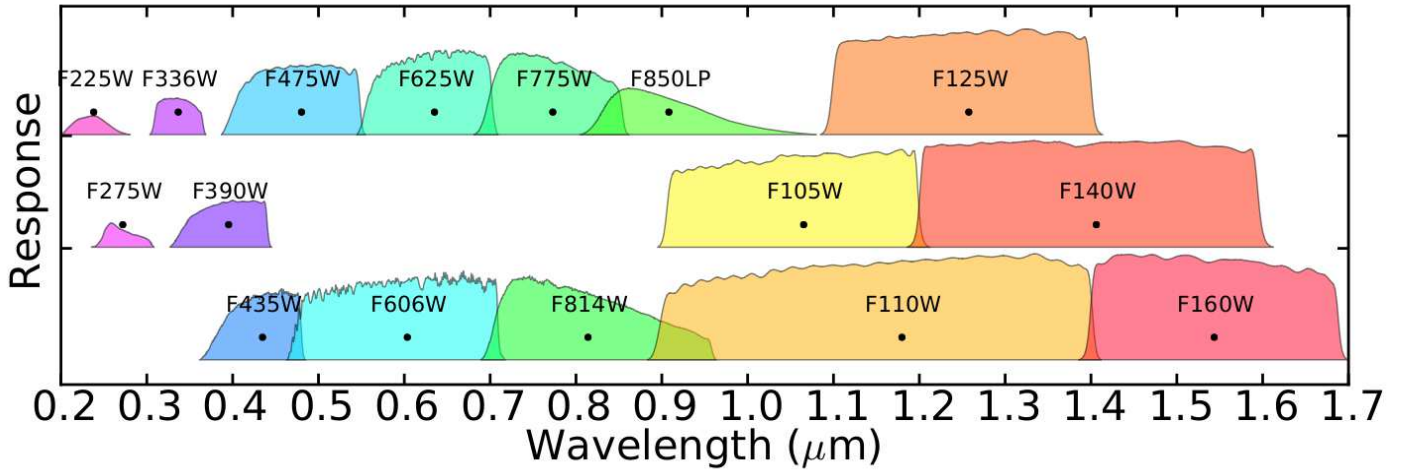}
\caption{The HST filter set used by CLASH, showing the transmission as a function of wavelength from UV to NIR.}
\label{fig:clashfilters}
\end{figure*}

\subsection{CLASH spectroscopic redshifts}
\label{subsec:spec-z}

We use a sample of spectroscopic redshifts (spec-z's) which were primarily acquired as part of our VIMOS/VLT Large 
Program 186.A-0798, targeting the 14 southern hemisphere CLASH clusters \citep{Rosati13} 
(for the dynamical mass profiles of MACS1206, see \citet{Lemze13} and \citet{Biviano13}).  This programme 
targets many thousands of galaxies around CLASH clusters drawn from our wide-field Subaru SuprimeCAM imaging data. 
In this paper, we use the 689 galaxies that have a spectroscopic redshift at least 80\% secure 
and that also lie within the narrower CLASH/HST field of view that
contains our 16-band imaging data. In addition, we have  included spec-z's from the literature and from observations at
the Magellan Telescope and the MMT. 

\begin{figure}[!h]
\resizebox{\hsize}{!}{\includegraphics{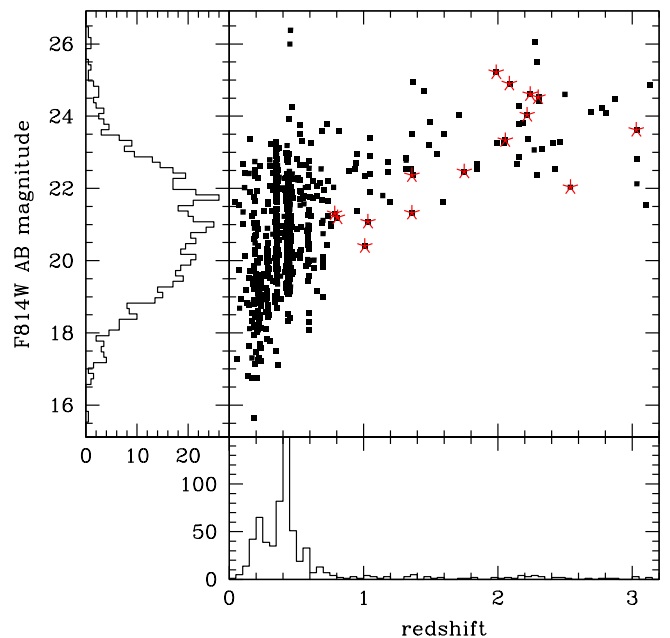}}
\caption{The spec-z sample which we use to assess the photo-z quality. Magnitude (F814W) and spectroscopic redshift 
from 17 cluster fields. Most spec-z are either
lensed galaxies at $z_s>1$, whereas galaxy cluster members at $z_s<1$. The red stars show the obvious strong-lensing
arcs for which we have a spectroscopic redshift.}
\label{fig:mz_spec-z}
\end{figure}

The spec-z's span a magnitude range of 18$<$m(F814W)$<$26 as shown in \autoref{fig:mz_spec-z}. 
The spec-z sample has a higher fraction of red sequence galaxies than a randomly chosen
galaxy sample since one of the VLT programme's primary objectives is cluster dynamics.
Lensed arcs have also been targeted but they do not represent a large fraction 
of the spec-z sample. 
We use the spec-z sample only to assess the photo-z
accuracy but do not use it for spectral template calibration.  

Following \citet{Maturi13}, we
divide the spec-z sample into two: a low redshift
sample containing 584 galaxy cluster members
and foreground galaxies ($z_s\leq z_{cluster}+0.2$), and a high-redshift sample
containing 106 lensed background galaxies ($z_s>z_{cluster}+0.2$).
In both samples, we only use redshifts considered to be at least 80\% secure.

\section{Photometric redshift estimation}
\label{sec:code}

As mentioned above, we make use of two template-based photo-z codes: Le Phare and BPZ. In this section, we briefly describe
these codes as well as their validity for application to the CLASH/HST data.

\subsection{Le Phare}
Le Phare\footnote{www.cfht.hawaii.edu/$\sim$arnouts/lephare.html} 
is a public photo-z code based on a template fitting method used in \citet{Ilbert06,Ilbert09}. 
It uses a $\chi^2$ minimization defined as :
\begin{equation} \label{eq:chi2}
\chi_ {model}^{2}=\sum_{i=1}^{n}([F_{obs}^{i}-\alpha F_{model}^{i}]/\sigma^{i})^{2},
\end{equation}
where $F_{obs}^{i}$~and $F_{model}^{i}$ are the measured and the template
model fluxes in filter $i$ and $\sigma^{i}$ is the corresponding
measured flux error. We use the templates optimized
for the COSMOS photo-z library as described in \citet{Ilbert09}. This library is composed 
of 31 templates ranging from ellipticals to starburst types. There are three elliptical galaxy templates
and seven spiral galaxy templates generated by \citet{Polletta07}.
In addition, there are 12 templates of starburst galaxies from \citet{BC03} with ages between 0.03 
and 3 Gyr. 
For the intermediate and late-type populations, we apply extinction law  
from \citet{Calzetti00} with extinctions values of E(B-V)=0.1,0.2,0.3,0.4,0.5.
Le Phare does not interpolate between templates.
We add emission line fluxes to the magnitudes derived in the template library. This is
a standard procedure from Le Phare aiming to better reproduce the colours 
of galaxies with significant ongoing star formation, as described in \citet{Ilbert09,Jouvel09}. 

\subsection{BPZ}
BPZ \footnote{http://www.its.caltech.edu/$\sim$coe/BPZ/} \citep{Benitez00,Benitez04,Coe06}, 
is a Bayesian photo-z estimation based on a template fitting method.
BPZ uses a new library \citep{Benitez13} composed by five SED templates originally drawn from PEGASE 
\citep{Fioc97} that were adjusted and re-calibrated from FIREWORKS photometry and spectroscopic redshifts
\citep{Wuyts08} to optimize its performance. 
The FIREWORKS data includes photometry for galaxies down to 24.3 (5 $\sigma$ K band)
out to $z\sim3.7$ from 0.38 to 24 $\mu$m. In analyses of large data sets with high quality
spectra, this template set yields less than 1\% outliers, demonstrating that it
emcompasses the range of metallicities, extinctions, star formation histories observed
for the vast majority of galaxies. These templates include nebular emission lines as 
implemented by \citet{Fioc97}. 
In addition, the BPZ library also includes five templates from GRASIL and a starburst galaxy.
BPZ library is then composed by 11 galaxy templates and different sets of priors based on redshift, types, and
magnitude distribution from existing surveys and allows for interpolation between templates
in the $\chi^2$ minimization.

BPZ is the default method used in the pipeline to derive photo-z for CLASH.

\subsection{Definition of photo-z quality}
 \label{subsubsec:photo-zq}
To assess the quality of the CLASH photo-z, we compute the scatter, mean, 
and outliers number of the photo-z deviation, $\Delta z_p = z_p -z_s$.
The scatter is measured by the normalized median absolute deviation (NMAD) 
described in \citet{Hoaglin83}, which is given by $1.48*median[\Delta z_p/(1+z_s)]$. 
This is a robust estimator, less liable to biasing by catastrophic outliers than the sample standard deviation.
The mean ($\mu$) is the mean value of $\Delta z_p/(1+z_s)$ and the number of outliers ($\eta$) 
is defined as galaxies whose $|\Delta z_p|/(1+z_s)>0.15$.

We select sub-samples of the photo-z based on estimates of the reliability of each photo-z (called
$odds$ for BPZ and $pdz\_best$ for Le Phare). This parameter is
the integral around the best-fit redshift of the probability distribution of the redshift $p(z)$
given by the $\chi^2$ fitting. 
We take a window of 2.5 times the photo-z precision expected from mock catalogue 
for the CLASH survey which is a window of 0.05 around the best-fit redshift.
The $odds$ is a number between 0 and 1 which is
equal to 1 if the $p(z)$ has a single, well-defined narrow peak, while it will be at lower value 
if there are multiple peaks or if the peak is broader than 0.05 in redshift.

\subsection{Template Library Representation}

To derive good photo-z's one requires a template library that contains spectra that are representative 
of the galaxy types and colours observed by CLASH. The BPZ and Le Phare 
redshift estimates are based on $\chi^2$ minimizations respectively with and 
without the help of sets of priors and template combination. 

\begin{figure}[!h]
\resizebox{\hsize}{!}{\includegraphics{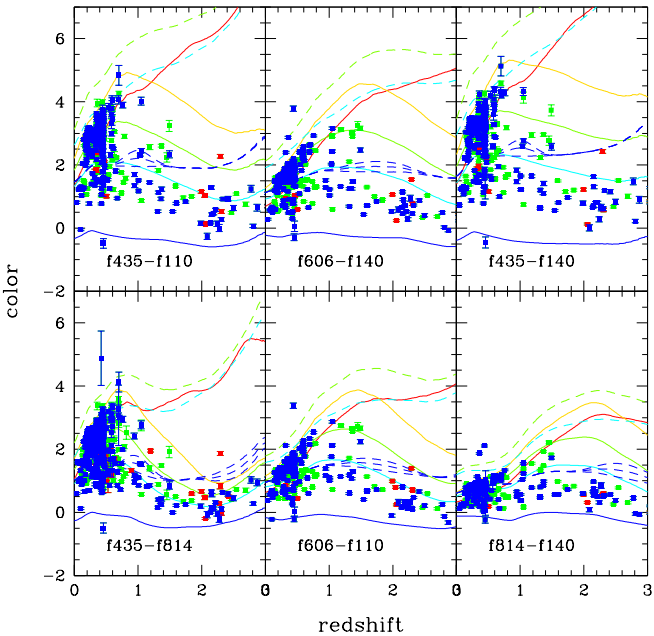}}
\caption{HST Color as a function of galaxy redshift. 
The solid and dashed lines show the COSMOS library colours intrinsic and extended 
using Calzetti extinction law. While the library is composed with 31 models, we show two ellipticals (red-gold), 
one intermediates (green) and two starbursts (cyan-blue). The blue, green and 
red squares correspond to the observed colours of the 
spectroscopic redshift sample. The blue squares are the galaxies with good photometric redshift
2\%(1+z), the red squares are catastrophic 
outliers ($>15\%(1+z)$) for which $odds$ are higher than 90\%. 
The green squares are the rest of the sample. The photo-z library colours match the
colour of CLASH observations well.} 
\label{fig:colcol_spec-z}
\end{figure}

The CLASH spec-z sample of 689 galaxies contains mainly galaxies with $z_s<1$. In this range, the strongest
colour gradient, from the Balmer/D4000 breaks, will happen in between F435W and F814W.  
\autoref{fig:colcol_spec-z} shows different colours as a function of 
redshift. The COSMOS library is represented
by the solid and dashed lines and the CLASH/HST spec-z sample is represented by the dots.
It is clear that the library provides good coverage of the spec-z sample.


\section{CLASH photo-z accuracy}
\label{sec:accuracy}

In this section, we study the photo-z accuracy and number of outliers for the galaxies
in the foreground and cluster members, for the lensed background galaxies, and the full spec-z sample
as explained in section \ref{subsec:spec-z}.
First, we study the overall accuracy for all 16 filters with Le Phare and BPZ then compare
different sub-sets of filters. 
Next, we investigate photo-z accuracy based 
on fixed-aperture and the automated Kron photometry ($mag\_auto$) and, finally, we discuss 
the photo-z performance in an idealized scenario based on mock catalogues.

\subsection{Photometric redshift comparison: BPZ and Le Phare}
\label{subsec:bpz_lpz}
Now we compare the BPZ and Le Phare photo-z estimates of
the spec-z sample.
\autoref{fig:photoz_flags} shows the photo-z vs
spec-z for the full spec-z sample.
For Le Phare photo-z estimation, we include the zeropoints or 
systematic shifts in the magnitudes. 
Table \ref{tab:shifts} shows the shift's values that have been 
derived using the spec-z sample up to magnitude m(F775W)$<$24.  
These shifts account for the mismatches
between the CLASH colours and the photo-z library colours and possible problems, if any, 
on our transmission (filter-mirror-detector) throughput model. The shifts are less than a tenth of a magnitude for 
the optical and NIR filters and can go up to 0.4 mag  for the UVIS filters. 
The variance of the zeropoints in the UVIS shifts is large, especially for the filters F225W and F275W.
The spec-z sample is dominated by galaxies at $z<0.8$ (90\%) and there is either a weak detection or no detection 
in the UVIS filters.
In the case of the UVIS filters, incompleteness in the template libraries is expected to
be the main contributor to the variance in the zeropoint estimation, in particular the reduced number of star-forming
galaxy templates. In addition, interpolation between the templates is not
allowed in Le Phare. Dust extinction also adds degeneracies to the zeropoint computation.
These last factors make the zeropoint computation in the NUV wavelenghts more challenging.
  
\subsubsection{Photometric redshift results}
\begin{figure}[!h]
\resizebox{\hsize}{!}{\includegraphics{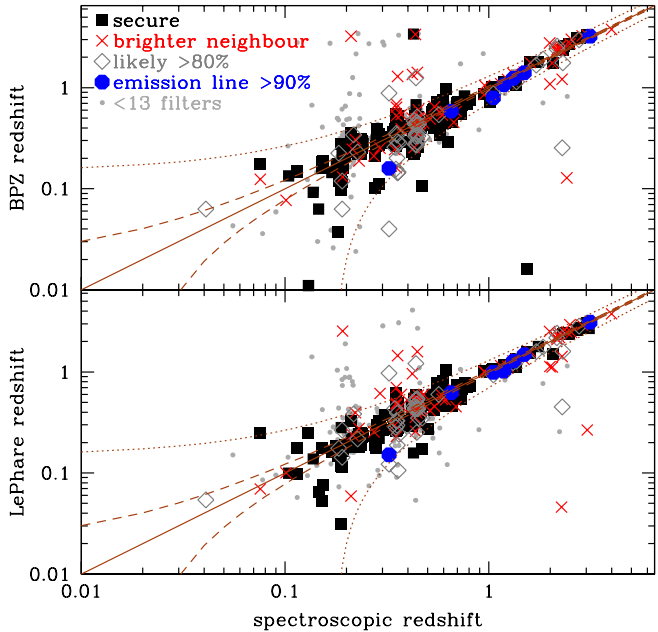}}
\caption{Spectroscopic vs photometric redshifts for the CLASH spec-z sample showing
Le Phare (bottom) and BPZ (top) performances for the spec-z secure redshift
in black squares, likely redshift in dark grey diamonds and one emission line
redshift in blue hexagons observed in at least 13 filters. 
We also added galaxies with a number a dectection in less than
13 filters in light grey dots. The red crosses mark the spec-z for which the closest HST photometry
match had a possible other solution at less than 3$"$. The brown dashed, and dotted curves correspond to
$z_p=z_s\pm0.02(1+z_s)$, and $z_p=z_s\pm0.15(1+z_s)$.} 
\label{fig:photoz_flags}
\end{figure}

\begin{figure}[!h]
\resizebox{\hsize}{!}{\includegraphics{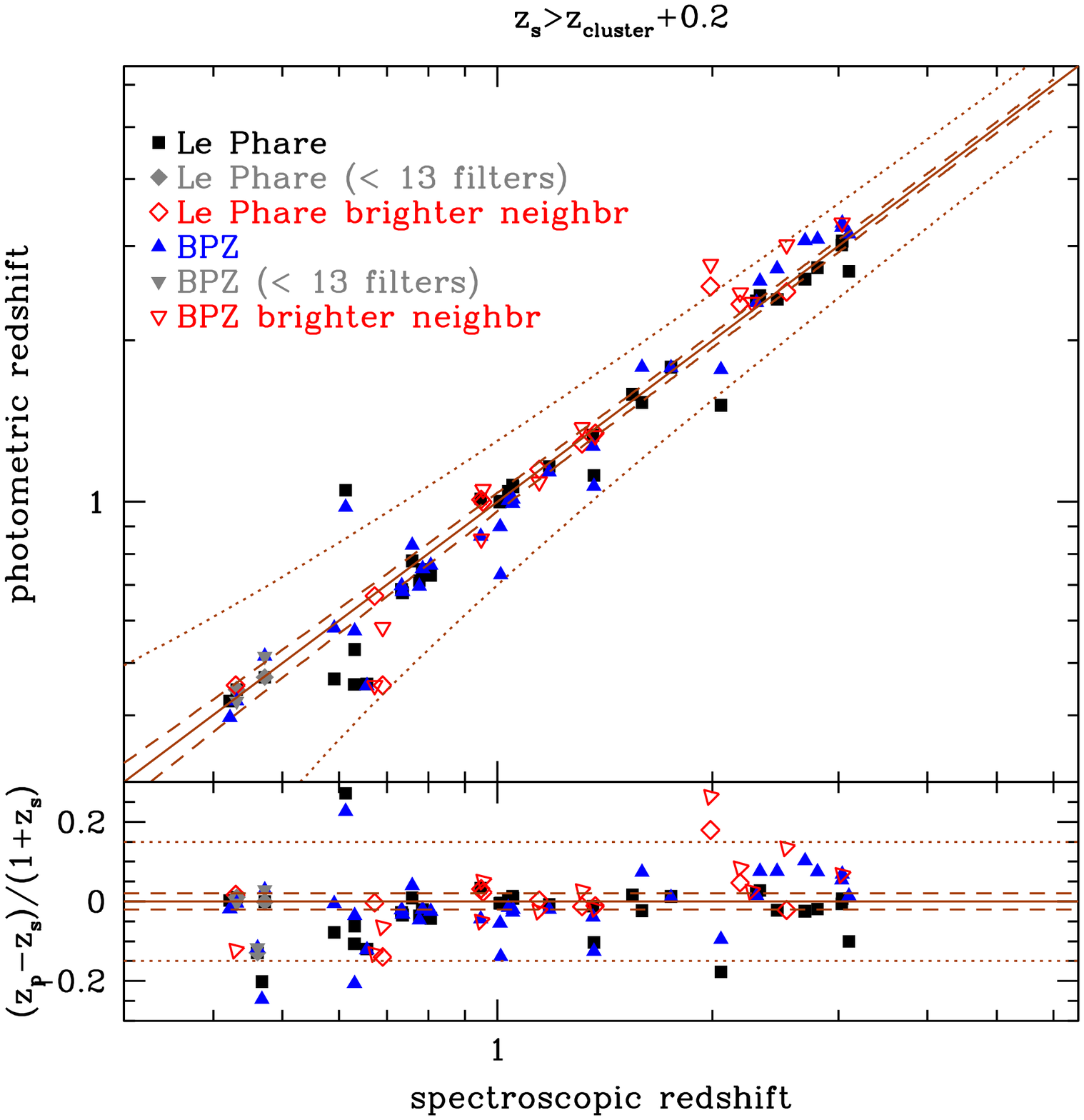}}
\resizebox{\hsize}{!}{\includegraphics{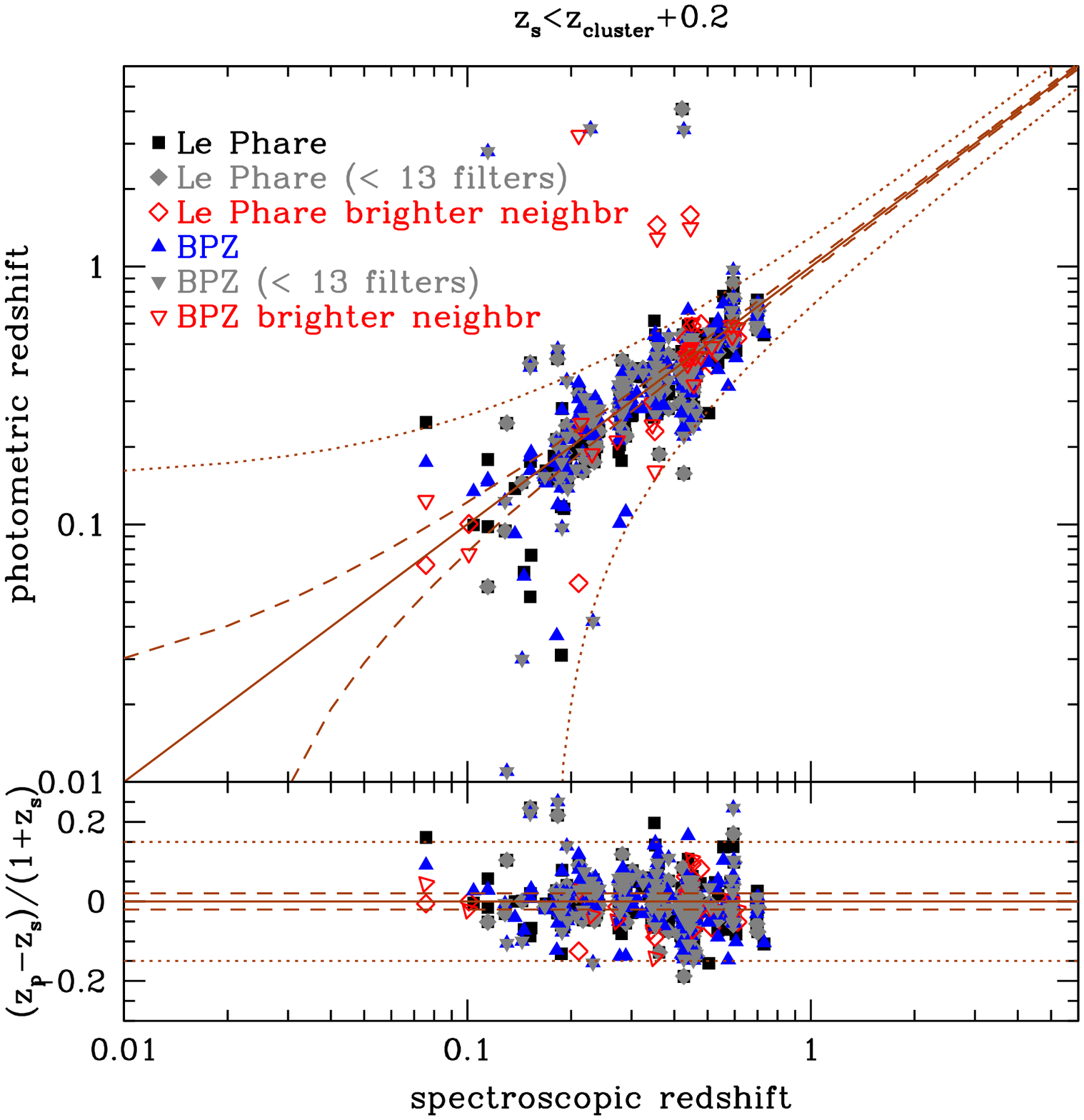}}
\caption{Spectroscopic vs photometric redshifts for the CLASH spec-z sample
whose magnitude and redshift distributions are shown in \autoref{fig:mz_spec-z}.
The black square and blue triangles are the BPZ and Le Phare photo-z
for galaxies observed in at least 7 bands with secure spec-z. The grey triangle and squares are
respectively BPZ and Le Phare secure redshifts observed in less than 13 filters. The red empty triangle
and square are the spec-z with a brighter neighbour within 3$"$ of the closest match to the spec-z. 
Top figure correspond to galaxy lensed images behind the
cluster at a redshift of $z_s>z_{cluster}+0.2$. Bottom figure correspond to
galaxy cluster member and foreground galaxies at $z_s\leq z_{cluster}+0.2$.}
\label{fig:zpzs_allclusters}
\end{figure}

\autoref{fig:photoz_flags} and \autoref{fig:zpzs_allclusters} show BPZ and Le Phare photo-z vs spec-z for galaxies observed 
in at least 13 filters.  Both methods show very good results for galaxies with secure spec-z's.  
There are only a few significant outliers (as defined in Section \ref{subsubsec:photo-zq}).  
Some of these have brighter galaxies (or galaxy segments) nearby (within 3$"$).  
Light from these brighter neighbors may have entered the spectroscopic slits in some cases.  
Relative astrometric uncertainties of ~1$"$ between the spectroscopic and HST catalogues may 
also contribute to occasional incorrect cross-matching.  
Of the remaining significant outliers, one is at the edge of the WFC3/IR field of view 
and another has a spec-z lacking a quality flag.  In futher analysis, we will only use  
the 554 secure spec-z (out of 689 total) to accurately assess the photo-z quality.

We choose to look at galaxies observed in at least 7 filters
to cover the optical wavelength range and with a S/N$>$10 in F775W. With the latter selections
we have a sample of 63 lensed background galaxies and 327 foreground-cluster galaxies. 
On these galaxy samples, BPZ and Le Phare have good performances for foreground and cluster member galaxies 
with an NMAD of 3 to 4\%(1+z), respectively, with and without 
photo-z quality cuts. For the lensed sample, we achieve a precision of 2.4 to 3\%
for Le Phare when using the photo-z estimations which include systematic shifts.
Without systematic shifts, BPZ and Le Phare have similar photo-z scatter 
with an NMAD of 5\%(1+z). 
\autoref{fig:zpzs_allclusters} show Le Phare photo-z with systematic shifts and BPZ
for lensed background galaxies and foreground-cluster galaxies. 
Over the whole redshift range, both Le Phare and BPZ have separately 18 catastrophic outliers. 
One could remove 7 out of the 18 outliers in discarding galaxies which have different values for Le Phare and BPZ photo-z. 
For the 11 other outliers, both BPZ and Le Phare have very close photo-z estimations. 
As explained in the latter section, catastrophic outliers are either galaxies observed in less than
13 filters, unsecure spec-z or a mismatch of the spec-z. We note that the source background
galaxies have less outliers than the foreground and cluster galaxies. 
Most galaxies living inside clusters have red colours from their old star population. 
Photo-z libraries do not include the reddest galaxies such as the BCG.
To improve further our photo-z estimations, there are ongoing efforts to create better templates for galaxies in cluster
environments \citep[e.g.][]{Greisel13}. Including these models would improve   
the photo-z for cluster members.
In section \ref{sec:arcs}, we study the photometry and
photo-z of the lensed images for one CLASH cluster, using customized photometry to have more 
control on the aperture shape in order to achieve a better S/N and 
have more accurate colours for the photo-z estimations. 

\subsubsection{Photometric redshift quality estimators}

\begin{figure}[!h]
\resizebox{\hsize}{!}{\includegraphics{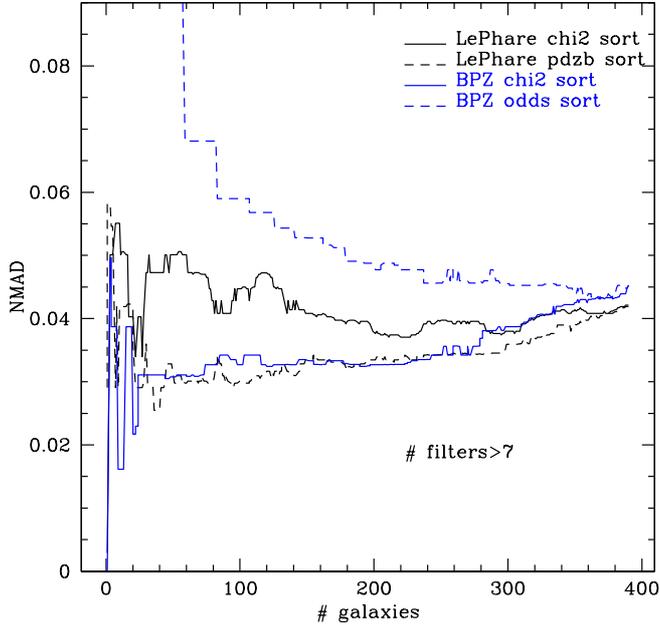}}
\caption{Photoz precision for the CLASH spec-z sample as a function of 
the number of galaxies sorted using
BPZ $odds$/Le Phare $pdz\_best$ (dashed lines) or $\chi^2$ (solid lines) from the photoz fit.
Blue and black lines show respectively BPZ and Le Phare results on the sample
of galaxies observed in at least 7 filters.}
\label{fig:nmad_nbrgal}
\end{figure}

In this section, we study the parameters which gives an estimation of 
the photo-z quality such as the $\chi^2$ and $odds$/$pdz\_best$
defined in section \ref{sec:code}, the 1 and 2 $\sigma$ confidence region
of the best-fit redshift. 
\autoref{fig:nmad_nbrgal} show the photo-z precision as a function 
of the number of galaxies sorted using
$\chi^2$ or $odds$/$pdz\_best$. According to \autoref{fig:nmad_nbrgal}, 
BPZ $\chi^2$ and Le Phare $pdz\_best$ 
give a good estimation of the photo-z quality for their respective codes. 

In order to study the reliability of the photo-z estimation,
we look at the 1- and 2-$\sigma$ confidence region. We compute
the confidence region using the cumulative values of $|z_p-z_s|$.
We find 62\% of galaxies with $z_p$ within
1-$\sigma$ of $z_s$ for Le Phare. 
We find respectively 96\% and 90\% of $z_p$ within 2-$\sigma$ of $z_s$
for Le Phare and BPZ. To assess the reliability of these estimations,
we bootstrap the 670 galaxies of our sample and find standard deviations of
0.61$\pm$0.02 and 0.956$\pm$0.008 for respectively the 1 and 2-$\sigma$
confidence region of Le Phare. We find 0.906$\pm$0.011 for BPZ
2-$\sigma$ confidence region. However, these calculations
assume that p(z) is symmetric. We looked at $z_p-z_s$
and find 40\% and 82\% of $z_s$ within the 1 and 2-$\sigma$
Le Phare confidence region and 58\% of $z_s$ within BPZ 2-$\sigma$
confidence region. 
The 1 and 2-$\sigma$ regions are under-estimated for our spec-z sample.
Our spec-z sample is mainly composed of bright galaxies which belong
the galaxy clusters. The p(z) of bright galaxies are usually narrow and do
not follow a gaussian profile. This result is then not unexpected. We note this
could also come from double-peaked p(z) from the Lyman-Balmer break confusion. 
When using the photo-z estimates, one should be careful in using
a single best value photo-z (mean or median of the p(z)) and rather use the full p(z)
distribution. The p(z) captures the whole information output by photo-z codes
which help to decrease biaises.


\subsection{Photometric redshift and systematics from the photometry}
\label{subsec:zp_sys}

We select galaxies observed in at least  13 filters with a S/N$>$10 in
F775W without selection in magnitude or redshift. 
We do not select on the quality of the photo-z fits ($odds$), in order to have
identical galaxy samples for fair comparisons between sub-samples.
This resulting sub-sample contains 289 galaxies that represents 42\% of the spec-z
galaxy sample with full HST coverage.

We show the utility of all filters especially for the range of redshift for which we
want accurate photo-z. 
We compute photo-z for the six following sub-samples of filters :
\begin{itemize}
\item optical filters of the ACS camera
\item NUV+optical filters of the WFC3/UVIS+ACS cameras
\item ACS+NIR filters of the ACS+WFC3/IR camera
\item UVIS1(F390W)+ACS+NIR of ACS+WFC3
\item UVIS2(F336W+F390W)+ACS+NIR of ACS+WFC3
\item NUV+optical+NIR filters of ACS+WFC3 
\end{itemize}

Table \ref{tab:lpz_filters_spec-z} shows the scatter, median and number
of catastrophic outliers for the different
sub-samples of the CLASH filters mentioned above as defined in section \ref{subsubsec:photo-zq}.
The full filter set yields a scatter of 4.1\%(1+z), and 
8 ``catastrophic" outliers as defined in section \ref{subsubsec:photo-zq} (where
$|z_{phot} - z_{spec}|/(1+z_{spec}) > 0.15$).
Removing the two bluest filters of the UVIS camera (UVIS2+ACS+NIR) reduces
the scatter to 3.9\%(1+z). 
We do not expect the 2 UVIS filters to significantly improve photo-z accuracy
for galaxies at redshift lower than 0.8, only at redshifts greater than 0.8.
The bluest CLASH filters help
the photo-z determination for galaxies at z$>$0.8 when the Lyman break becomes visible
in the first UVIS filter F225W.

\begin{table}[!ht]
\caption{Characteristics of the photo-z distribution for 
sub-sets of the CLASH filters described in the above paragraphs. 
We use galaxies observed in at least 13 filters with a S/N$>$10 in F775W
resulting in 303 galaxies or 44\% of the spec-z sample.
The first column is the sub-set of filters in which we compute photo-z, the second and
third column are the NMAD and the median of the $z_p-z_s$ distribution. The last
column $\eta$ is the number of catastrophic outliers.}
\begin{tabular}{cccccccc} \hline\hline
Photometry & NMAD & $\mu[\Delta z_p/(1+zs)]$ & $\eta$\\
\hline\hline
"ACS" & 0.051 & -0.006 & 19 \\
"UVIS\_ACS" & 0.051 & -0.020 & 18 \\
"ACS\_NIR" & 0.044 & -0.009 & 16 \\
"UVIS1\_ACS\_NIR" & 0.041 & -0.013 & 9 \\
"UVIS2\_ACS\_NIR" & 0.039 & -0.010 & 8 \\
"UVIS\_ACS\_NIR" & 0.040 & -0.012 & 8 \\
\end{tabular}
\label{tab:lpz_filters_spec-z}
\end{table}
The photo-z of Table \ref{tab:lpz_filters_spec-z} include systematic shifts in the
photometry computed using the spec-z sample.
The shift's values are listed in Table \ref{tab:shifts}.

\subsection{Photometric redshift and aperture photometry}
\label{subsec:aper}
Now we study the impact of different choices of photometry, including fixed apertures and 
automated Kron photometry.

We consider fixed apertures with diameters ranging from 
2 pixels to 100 pixels (the pixel size is 0.065 arcsec in CLASH co-added images). 
We also measure Kron photometry \citep{Kron80,Bertin96}, based on ellipsoidal apertures whose shape is calculated
from the second moments of the light distribution. 
The Kron scaling factor was set 
to 2.5, and the minimum radius to be 3.3$\sigma_{iso}$ (the isophotal radius) which
is in the range of values suggested by the SExtractor manual. 
We compare with the photo-z based 
on the standard isophotal photometry using the entire spectroscopic redshift sample.

\begin{figure}[!h]
\resizebox{\hsize}{!}{\includegraphics{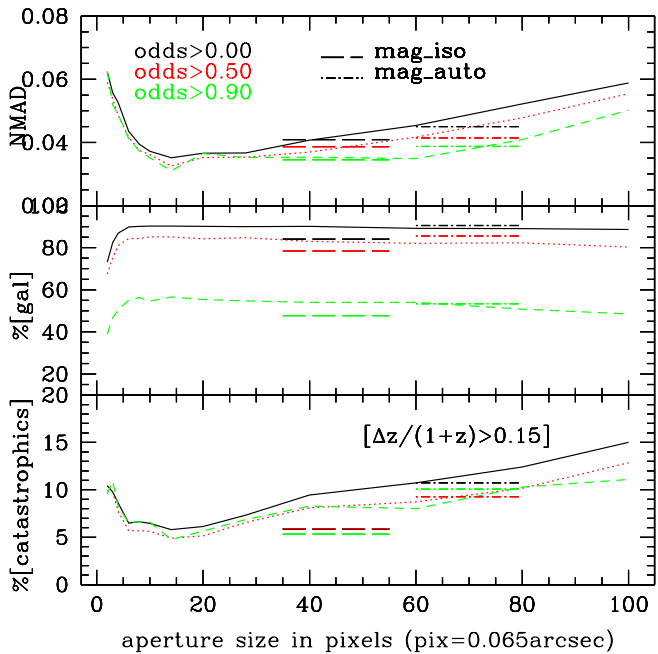}}
\caption{Top panel shows the photo-z scatter, middle panel is the percentage of galaxies selected 
and bottom panel is the percentage of catastrophic outliers for various choices of photometric apertures,
using the full spec-z sample. The black solid, red dotted, green long-dashed lines
show the results at $pdz\_best$ cut of 0,0.5,0.9. 
The optimal fixed aperture size is around 10-30 pixels. Isophotal
photometry (long-dashed lines) performs close to equally well
this range of fixed apertures. Automated Kron apertures (dashed-dotted lines) are not competitive.}
\label{fig:nmad_aper}
\end{figure}

\autoref{fig:nmad_aper} top panel shows the photo-z scatter for the different apertures.
There is a range of aperture diameters where the colour is maximized and the error 
minimized and that range corresponds to roughly 10 to 30 pixels. 
Automated Kron apertures do not give very good results compared to a carefully chosen 
aperture photometry with a NMAD of $4.5\%(1+z)$ for the highest $pdz\_best$ cut. 
Standard isophotal photometry (threshold scale at 1$\sigma$) gives the best results with a NMAD 3.0\%(1+z). 
\autoref{fig:nmad_aper} middle panel shows the percentage of
galaxies that enters in the NMAD calculation of \autoref{fig:nmad_aper}
top panel. 
Comparing the top and middle panel of \autoref{fig:nmad_aper},
it is clear that the $pdz\_best$ cuts
do not improve the scatter and number of outliers significantly when using the best range of
aperture photometry. 
The bottom panel of \autoref{fig:nmad_aper} shows the percentage
of catastrophic outliers as a function of aperture diameter. 
Again, fixed apertures of 10-30 pixels perform the best, minimizing the number of catastrophic 
outliers, optimizing the number of galaxies used, and $pdz\_best$ cuts do not improve the results 
significantly. Isophotal photometry gives very close results to the best range of fixed apertures. 

\begin{figure}[!h]
\resizebox{\hsize}{!}{\includegraphics{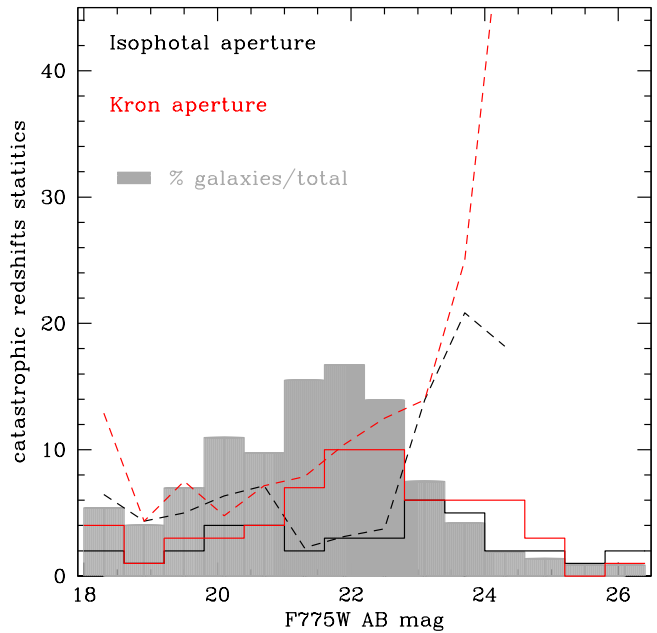}}
\caption{Number and percentage of catastrophic outliers as a function of F775W magnitude
in respectively solid and dashed lines. We find 39 and 64 catastrophic outliers for
respectively the isophotal and Kron aperture photometry. The grey shaded histogram shows the
percentage of galaxies compared to the entire spec-z sample in each magnitude bins.}
\label{fig:cata_mag_aper}
\end{figure}

\begin{figure}[!h]
\resizebox{\hsize}{!}{\includegraphics{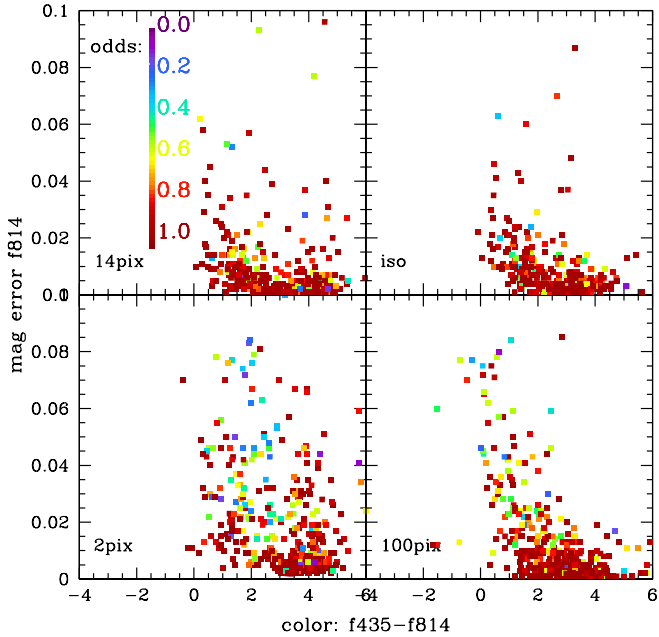}}
\caption{The magnitude error as a function of galaxy colour is shown for aperture photometry 
done using 2, 14, and 100 pixel apertures in the lower left, upper left, and lower right panels, 
respectively. The mag errors as a function of colour for isophotal photometry is shown in the upper right panel.
The 14 pixels aperture and isophotal photometry have higher colour accuracy than the 2
and 100 pixels aperture photometry which explains the photo-z results.}
\label{fig:colmag_aper}
\end{figure}
\autoref{fig:cata_mag_aper} shows the number and percentages of catastrophic outliers.
We compared both the isophotal and automated Kron photometry. The isophotal
photometry performs better than the Kron photometry with 39 catastrophic outliers 
compared to 64 for the Kron aperture using a catastrophic outliers definition defined in section
\ref{subsubsec:photo-zq}. It represents a percentage of 7\% for the isophotal photometry and
11\% for the Kron photometry. The percentage of catastrophic outliers 
by magnitude bins rises at magnitude m(F775W)$>$23. At m(F775W)$>$23 
the percentage of galaxies compared to the entire sample lowers as shown by the grey shaded histograms. 
The percentages are thus not very significant. 
\autoref{fig:colmag_aper} shows the F814W magnitude error as a function of
the colour F435W-F814W.
Bottom left, top left, bottom right panels show 
the colour -- photometric error relation for aperture diameters of 2, 14, and 100 pixels 
while the top right panel shows the corresponding isophotal photometry.
The isophotal and 14 pixels aperture photometry optimize the colour from the Balmer/D4000 
break while minimizing the photometric error.
In the case of a 2pixels and 100pixels photometry, the colour of 
the break is lost due to the lack of flux from small aperture size
or the high level noise due to the big aperture size. 
We choose to use isophotal photometry for our studies. 

We want to stress that these results have been computed using cluster galaxies.
These results might change for smaller starburst galaxies at higher redshifts.

\subsection{Photometric redshifts from simulated CLASH photometry}
\label{subsec:simul}

This section aims at testing the impact of systematics coming from galaxy properties 
on the photo-z estimations from the CLASH photometry in an ideal case.  
We use catalogues from Le Phare simulations that are described in \citet{Jouvel09}. 

We produce mock galaxy catalogues in the CLASH/HST bands based on a luminosity 
function that depends on both redshift and morphological type as derived from the GOODS survey
\citep{Giavalisco04,Dahlen05}.
The simulations do not include the background cluster light, characteristic of a cluster 
field, so this should be regarded as 'best-case' scenario.
The simulations are based on a mixture of templates
which includes elliptical galaxies up to starbursts types.
We miss the reddest galaxies such as the BCG's and the oldest galaxy population.
This would likely affect the photo-z accuracy of galaxies at $z_s<1$. 
It should however impact less the photo-z accuracy of the lensed 
background sample which have a higher fraction
of blue young galaxies. 

Input templates come from \citet{Coleman80}, extended in wavelength with synthetic spectra
from \citet{BC03}. There are five main templates and one starburst template which are
linearily interpolated to provide 66 independent characterizations of a galaxy's SED, which we refer to as the 
CE (Coleman Extended) templates. 
The luminosity function has been adapted to reproduce COSMOS colours,
number counts and redshift distribution as explained in \citet{Jouvel09}. 
The depth in each band is calculated to simulate the 
CLASH photometric depth shown in Table \ref{tab:depth}. The magnitude-error
relation follows a Gaussian distribution based on galaxy fluxes and does not 
take galaxy sizes into account. 

We produce four catalogues of increasing complexity in terms of galaxy physics.
Table \ref{tab:simul} summarizes the steps from simul1 (the simplest) to simul4 (most complex).
Simul2 includes extinction as described by \citet{Calzetti00}
for the late type galaxies, and simul3 also adds emission lines, using calibrations between
the star formation rate and emission line fluxes \citep{Kennicutt98}. For simul1-3 we use the same libraries to create the mock catalogues
and to calculate photo-z, i.e.~the photo-z library matches the truth. For
simul4 we use the same library as simul3 for the simulations, but the COSMOS library with
extinction and emission lines to calculate photo-z.

\begin{figure}[!ht]
\resizebox{\hsize}{!}{\includegraphics{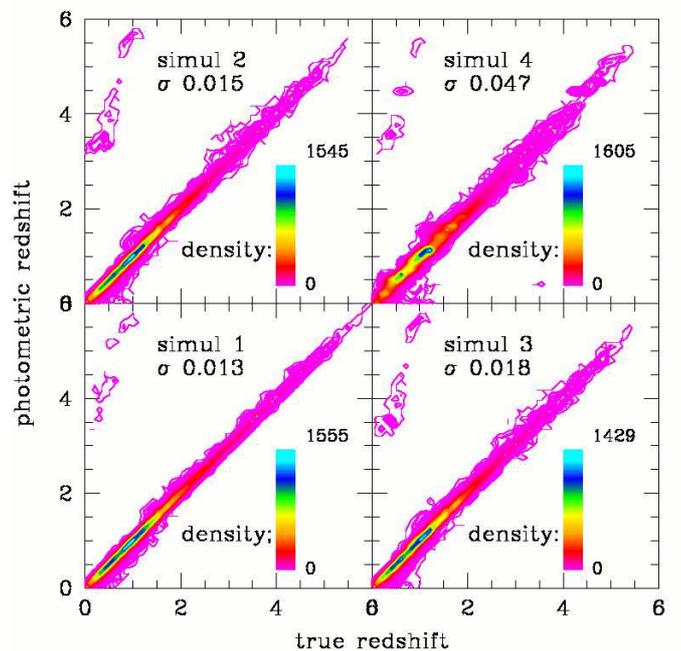}}
\caption{Photometric redshift results for simulations of the CLASH photometry using 
the depths listed in Table \ref{tab:depth}. From simul1 to simul4 we add complexity 
in galaxy physics such as extinction laws, emission lines and different galaxy libray as
explained in Table \ref{tab:simul} which increase the photoz scatter.}
\label{fig:zpzs_simul}
\end{figure}

\begin{table}[!ht]
\caption{Description of the simulations for which results are shown in \autoref{fig:zpzs_simul}.
Each row represents a different simulation and photo-z calculation, where the 2nd and 3rd column are the library
used in the mock catalogue construction and the photo-z library, respectively. The +ext and +eml
stands for the addition of extinction and emission lines to the libraries.}
\begin{tabular}{cccccccc} \hline\hline
X  & mock & photo-z\\
\hline\hline
simul 1 & CE & CE \\
simul 2 & CE+ext & CE+ext\\
simul 3 & CE+ext+eml& CE+ext+eml \\
simul 4 & CE+ext+eml & COSMOS syss\\
\end{tabular}
\label{tab:simul}
\end{table}

\autoref{fig:zpzs_simul} shows the photo-z results obtained from the
four catalogues. Again, the scatter is given by the NMAD.
The bottom left panel
is the simplest case in which we reach a precision of
0.013(1+z). Adding extinction (simul2) and emission lines (simul3) decreases the precision slightly.
Then, going to simul4 in the top right panel increases the scatter to 0.047(1+z).
For simul4 we computed the systematic shifts to do a first order correction
on the simulation and photo-z library colour differences.
In the best case, if the colours of our library are representative of
observations, we should be close to simul3 with a scatter of 0.018*(1+z).
We note that in the magnitude range of the arcs sample 24$<$m(F814W)$<$26, 
the NMAD in simul3 increase to 0.025(1+z).
We do not expect the library colours to be fully representative of the 
variety we observe in the galaxy population. A precision of 3\% to 4\%(1+z)
is within what we can expect from the CLASH observations.

Table \ref{tab:simul_photoz} shows the photo-z scatter and fraction of outliers
for the four simulated cases at redshifts $z_s<1$ and $z_s>1$. Using the CLASH data,
we reach a precision of $3.6\%(1+z_s)$ for galaxies at $z_s<1$.
This precision lies in between the 2\% precision achieved with an optimal fully 
representative template library (simul3) and the 5\% precision achieved with a non-optimized library (simul4).
At high-redshift $z_s>1$, the CLASH data yields a photo-z precision of
$2.6\%(1+z_s)$ and $3.1\%(1+z_s)$, respectively, with or without an $pdz\_best$ cut. 
Simul3 and simul4 show an NMAD of $1.7\%(1+z_s)$
to $4.3\%(1+z_s)$. We conclude that the actual performance of the CLASH photo-z are within expectations.

\begin{table}[!ht]
\caption{Photoz results from the four simulations of the CLASH photometry. We show the
NMAD (scatter) for galaxies at $z_s<1$ and $z_s>1$. $f_{cata}$ is the outlier fraction in \%
following the same definiton as $\eta$, $\eta$ being the number of outliers.}
\begin{tabular}{cccccccc} \hline\hline
simul & nmad[$z_s<1$] & $f_{cata}$[$z_s<1$] & nmad[$z_s>1$] &  $f_{cata}$[$z_s>1$]\\
\hline\hline
simul1 &  0.015   &    0.4\% &  0.011  &  0.2\% \\
simul2 &  0.016   &    1.1\% &  0.014  &  0.4\% \\
simul3 &  0.020   &    1 \%  &  0.017  &  0.4\% \\
simul4 &  0.050   &    1.7\% &  0.043  &  0.4\% \\
\end{tabular}
\label{tab:simul_photoz}
\end{table}

\section{Photometric redshifts of lensed galaxy images - MACS1206 example}
\label{sec:arcs}
In the previous sections, we mainly focused on photo-z for galaxies in the cluster
field. In this section, we focus on the photo-z quality of the "background lensed galaxies (i.e., the arcs).
 
Strong lensing analysis incorporate the photo-z estimates of individual galaxies in the
lens modelling. 
Photo-z scatter can have an impact on the strong lensing mass model. 
For example, erroneous photo-z values can lead to
model-tensions because the positions of lensed galaxies with inaccurate redshifts will not be correctly reproduced even if
the model is largely based on arcs with reliable redshift values and the model is, thus, mostly reliable. 
We thus explore the degree to which the scatter in our photometric redshifts can bias the 
derived cluster mass profile parameters (e.g., central concentration). In this section, we focus, 
in particular, on the lensed background galaxies. The arcs that are projected near bright 
cluster galaxies may be particularly susceptible to higher photometric errors. 

\begin{figure}[!ht]
\resizebox{\hsize}{!}{\includegraphics{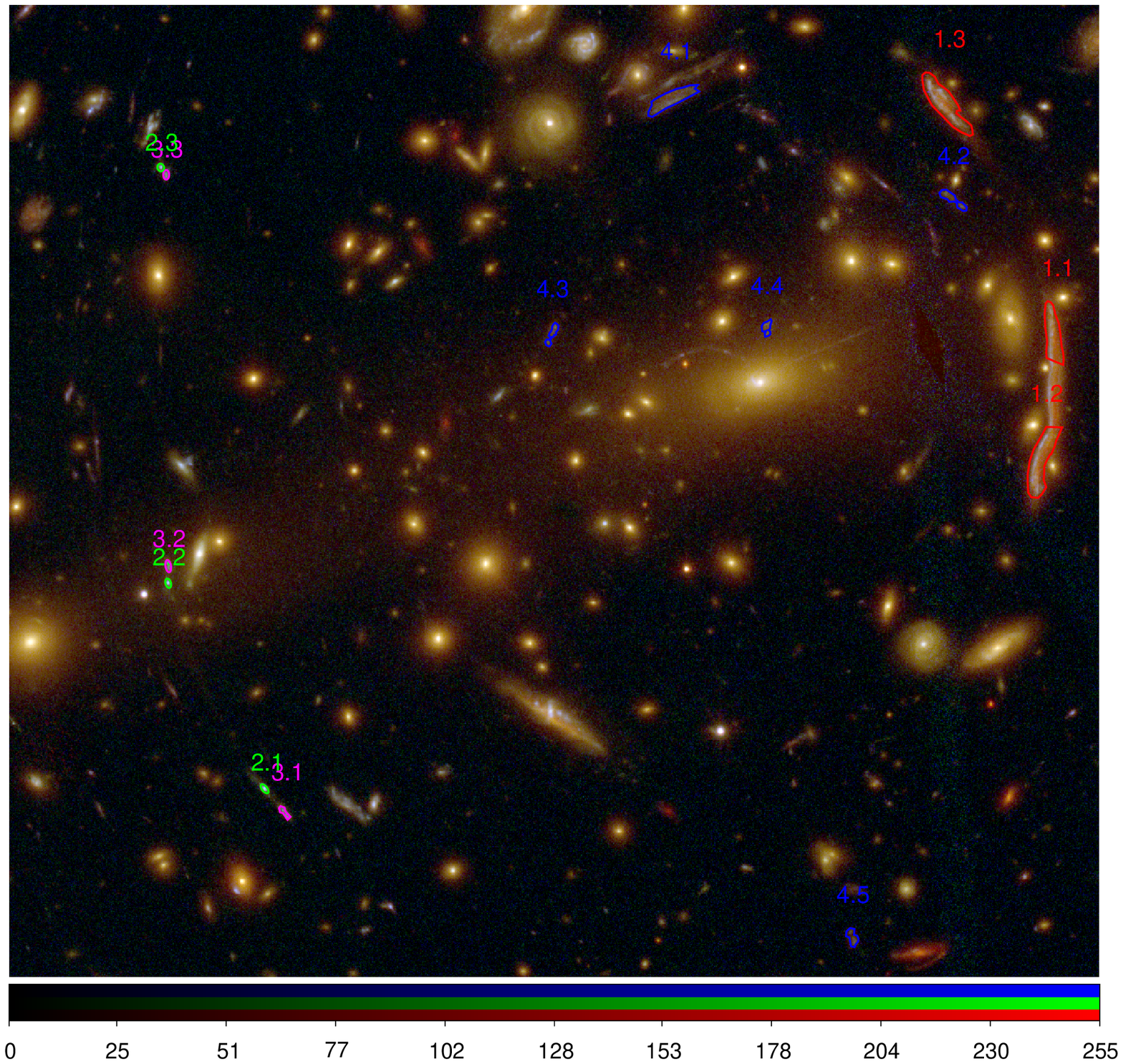}}
\caption{The galaxy cluster MACS1206 at z$\sim$0.44 in a combination of optical filters
from the CLASH/HST observations (blue= F435W + F475W, green = F606W + F625W + F775W + F814W + F850LP
red = F105W + F110W + F125W + F140W + F160W).
We show the strongly lensed multiple images 
described in \citet{Zitrin12} for which we have spec-z. 
Systems 1, 2, and 3 are highlighted in red, green, and blue, respectively.
The image segmentation in these cases has been defined manually to optimize the 
reliability of the arc photometry.
North is up, East is left. The pixel flux levels are given by the colourbar}
\label{fig:m1206_arcs}
\end{figure}

We study  in detail the photo-z measurements of fourteen lensed images, which have 
spec-z, behind the cluster MACS J1206.2-0847 (z=0.4385; MACS1206 hereafter).
This cluster was X-ray selected and
its lensing properties were first studied in the Massive Cluster Survey, MACS \citep{Ebeling01,Ebeling07,Ebeling10}.
MACS1206 has high X-ray luminosity and a relatively smooth X-ray surface brightness profile
which reflects the selection function of the MACS survey.
MACS1206 first mass model was derived from F814W/HST
imaging using a single giant arc in \citet{Ebeling09}.
Using the multi-band CLASH observations we have identified 47 new lensed images of 12 sources,
four of which have spectroscopic redshifts \citep{Zitrin12} as shown in 
\autoref{fig:m1206_arcs}.

The first system is the one used in \citet{Ebeling09} at a spectroscopic redshift of $z_s$=1.033.
The other systems were observed by the VIMOS/VLT Large Program 186.A-0798. Systems two and three are 
at $z_s$=3.03 and system four at $z_s$=2.54.

Below, we go through each system
and present photo-zs calculated with Le Phare and BPZ. We note that a first version
of the CLASH photo-z for the SL arcs of MACS1206 was presented in \citet{Zitrin12}. 
\citet{Zitrin12} had not included the F336W data and presented photo-z for the default SExtractor photometry. 
In this section, the photometry and photo-z of the MACS1206 arcs are studied in greater detail with
the SExtractor and tailored photometry.
Arcs and multiple images can have very complicated shapes, and the standard SExtractor
image detection and photometry may break the arcs into several parts, as shown by the segmentation
map in Appendix \ref{app:arcs}. This can affect the accuracy of the photometry for these objects.
In the following, we compare Le Phare and BPZ results based on both custom manual photometry
and on the standard SExtractor isophotal photometry for each arc.
In Appendix \ref{app:arcs}, we also show photo-z results from BPZ using the isophotal
photometry as well as the segmentation maps from SExtractor, cut-out images
of the strong-lensing arcs, and the probability distributions and best-fit
templates from BPZ.

\begin{figure*}[!ht]
\hbox{
\includegraphics[width=0.35\textwidth,height=0.3\textwidth]{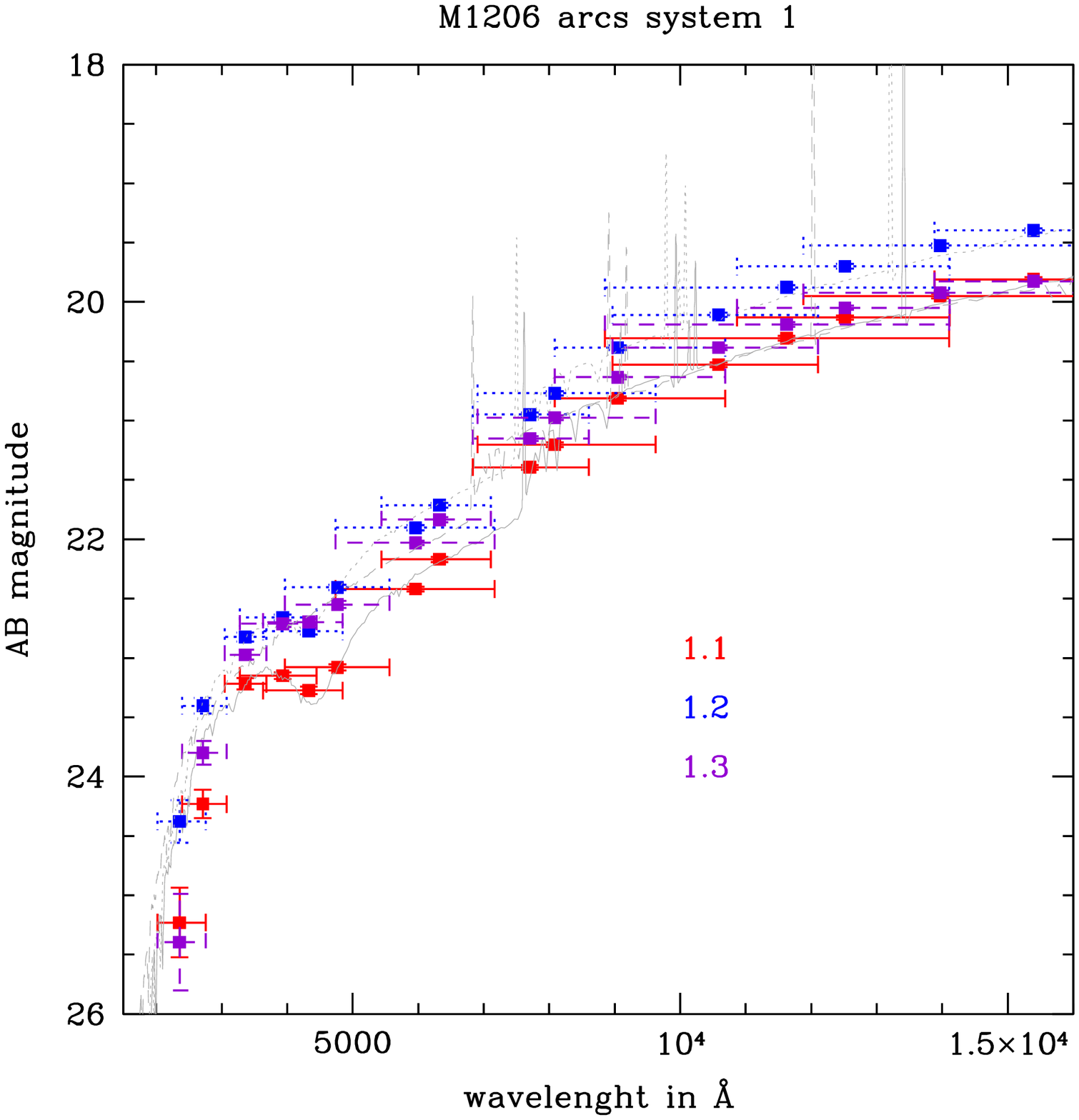}
\includegraphics[width=0.35\textwidth,height=0.3\textwidth]{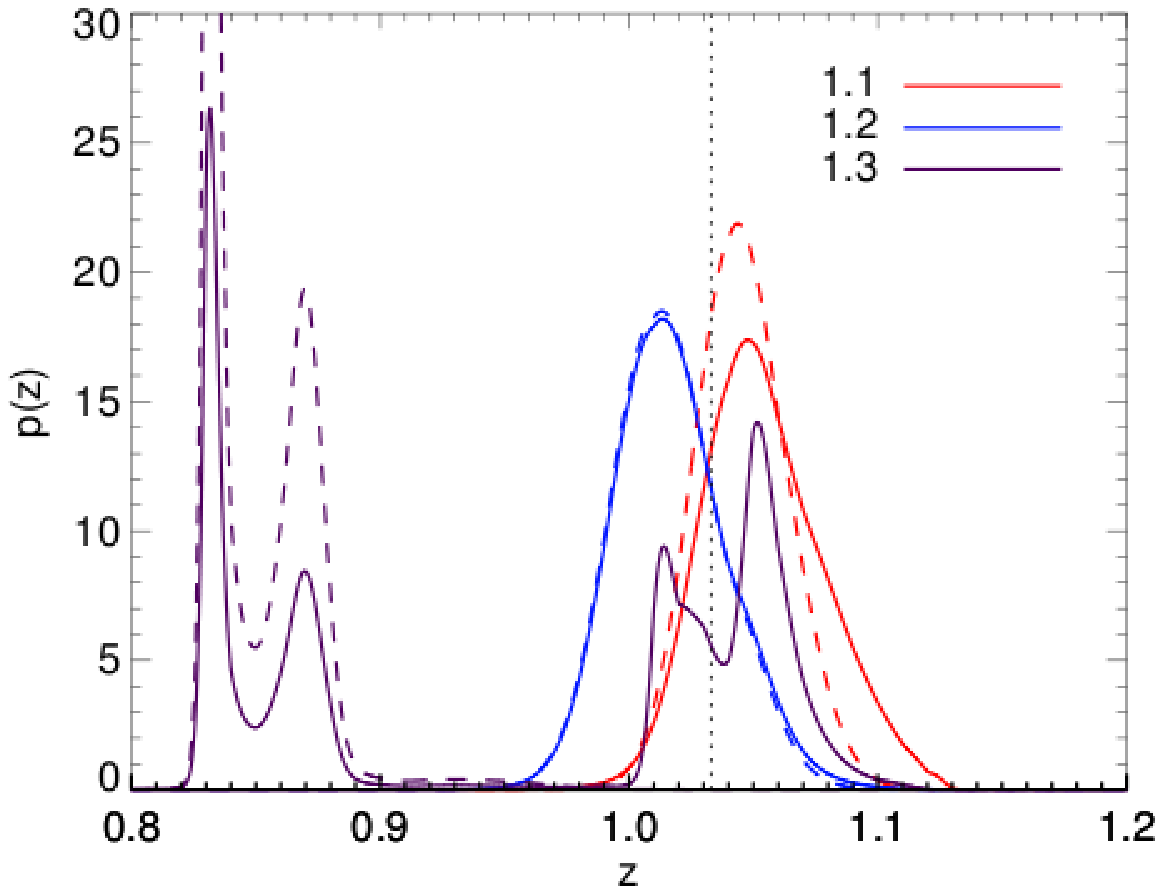}}
\hbox{
\includegraphics[width=0.35\textwidth,height=0.3\textwidth]{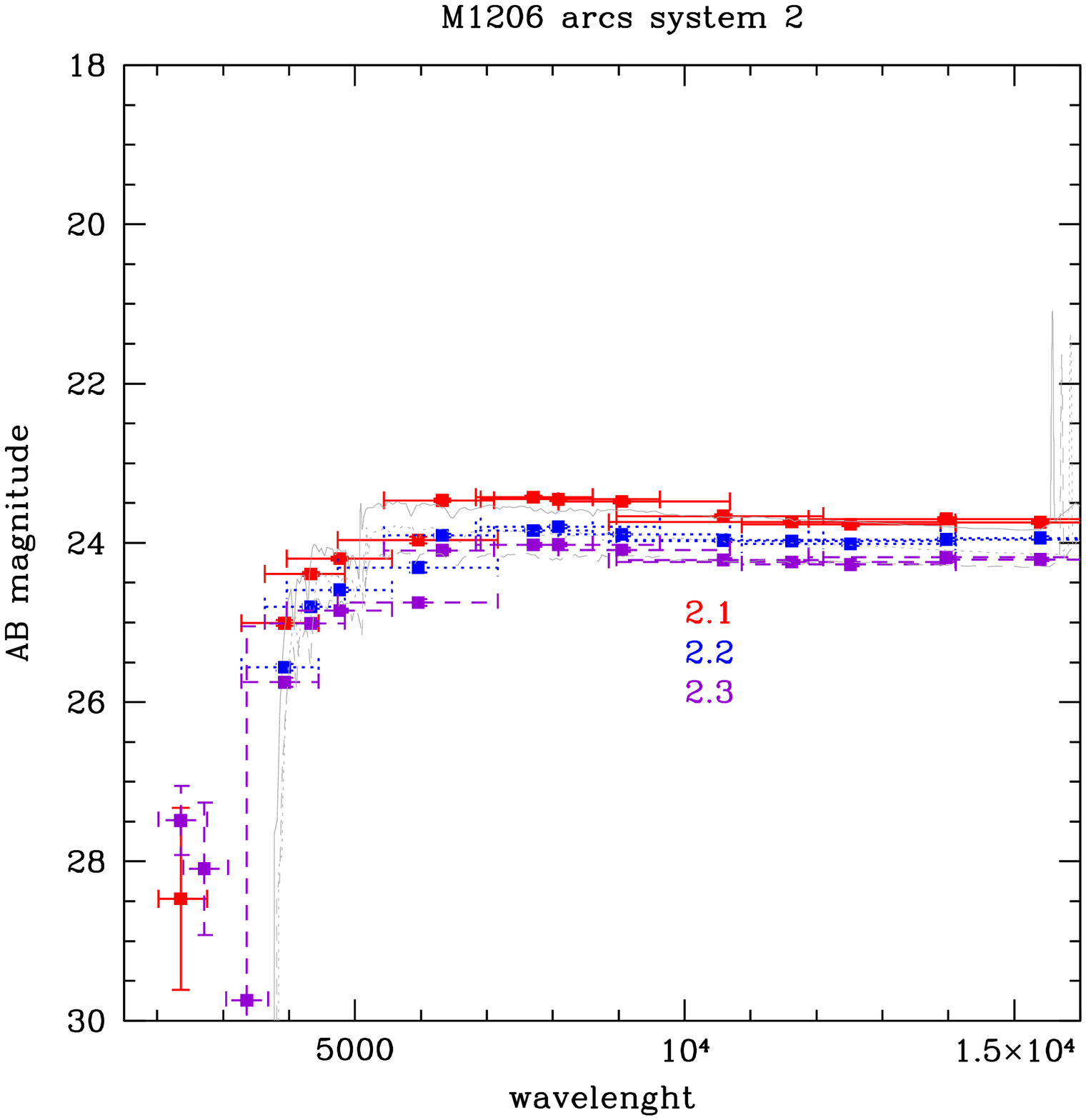}
\includegraphics[width=0.35\textwidth,height=0.3\textwidth]{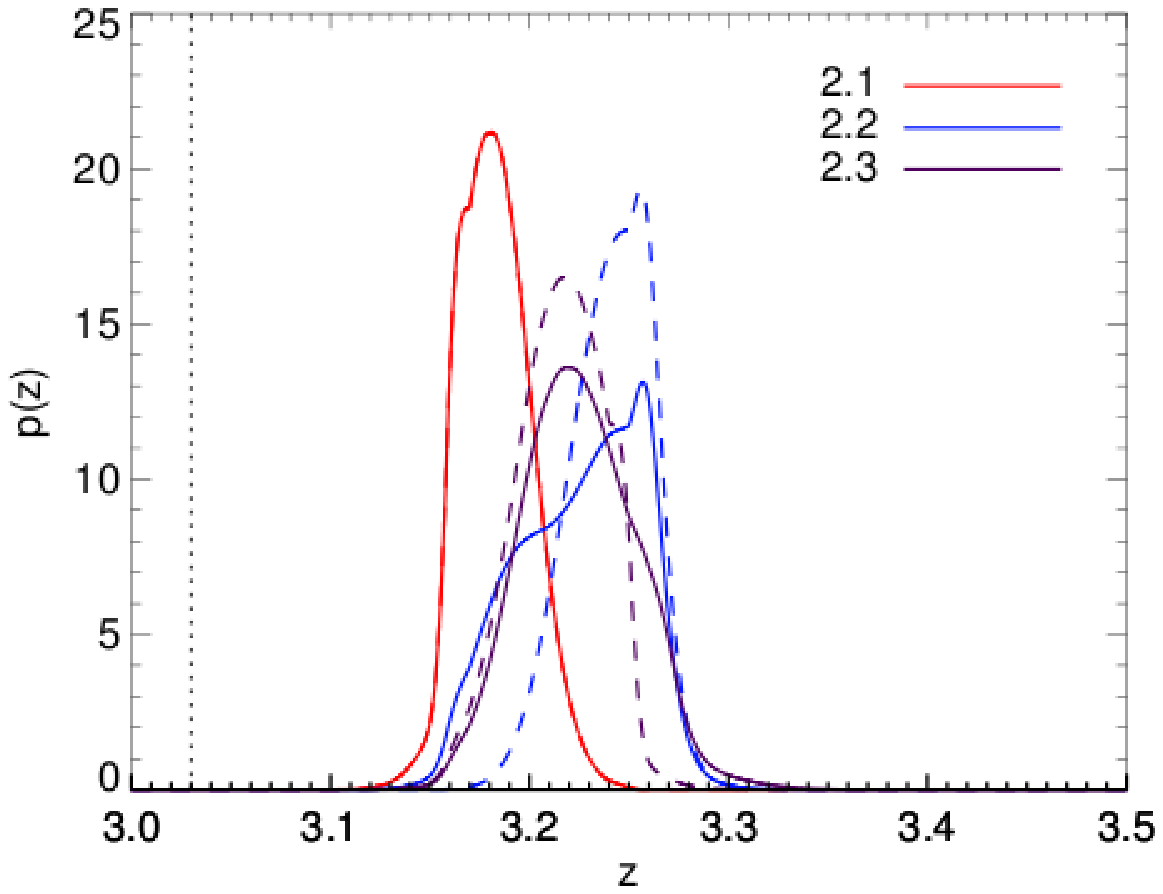}}
\hbox{
\includegraphics[width=0.35\textwidth,height=0.3\textwidth]{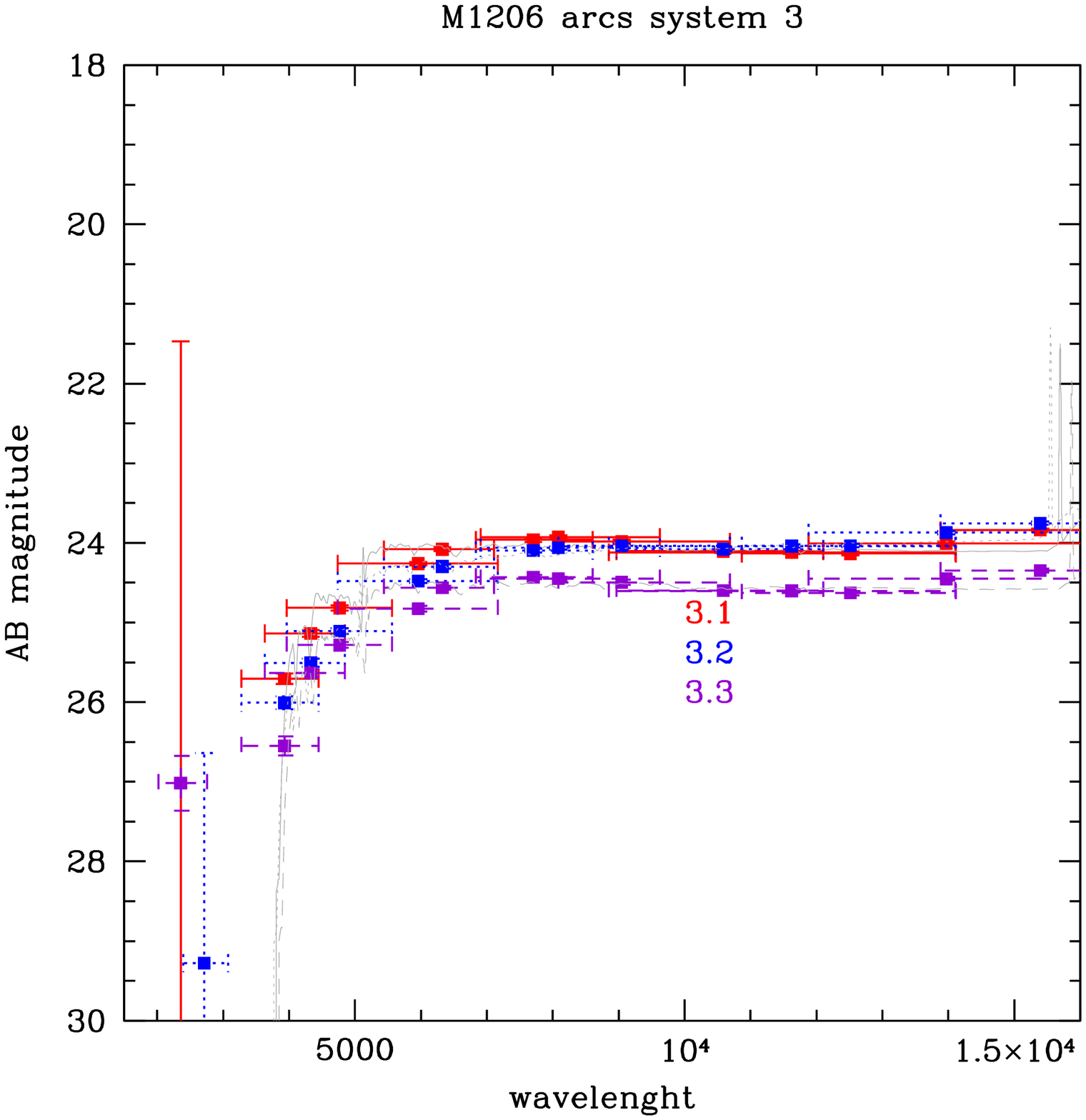}
\includegraphics[width=0.35\textwidth,height=0.3\textwidth]{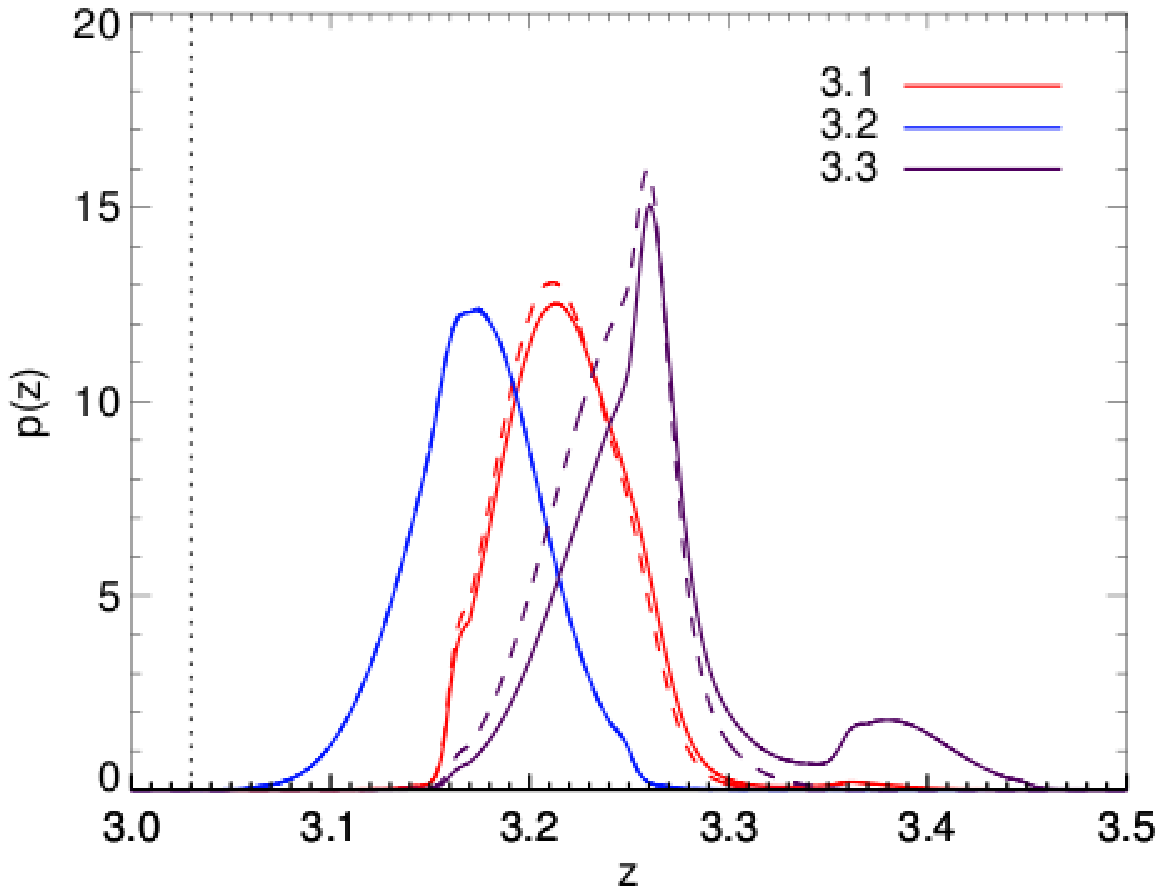}}
\hbox{
\includegraphics[width=0.35\textwidth,height=0.3\textwidth]{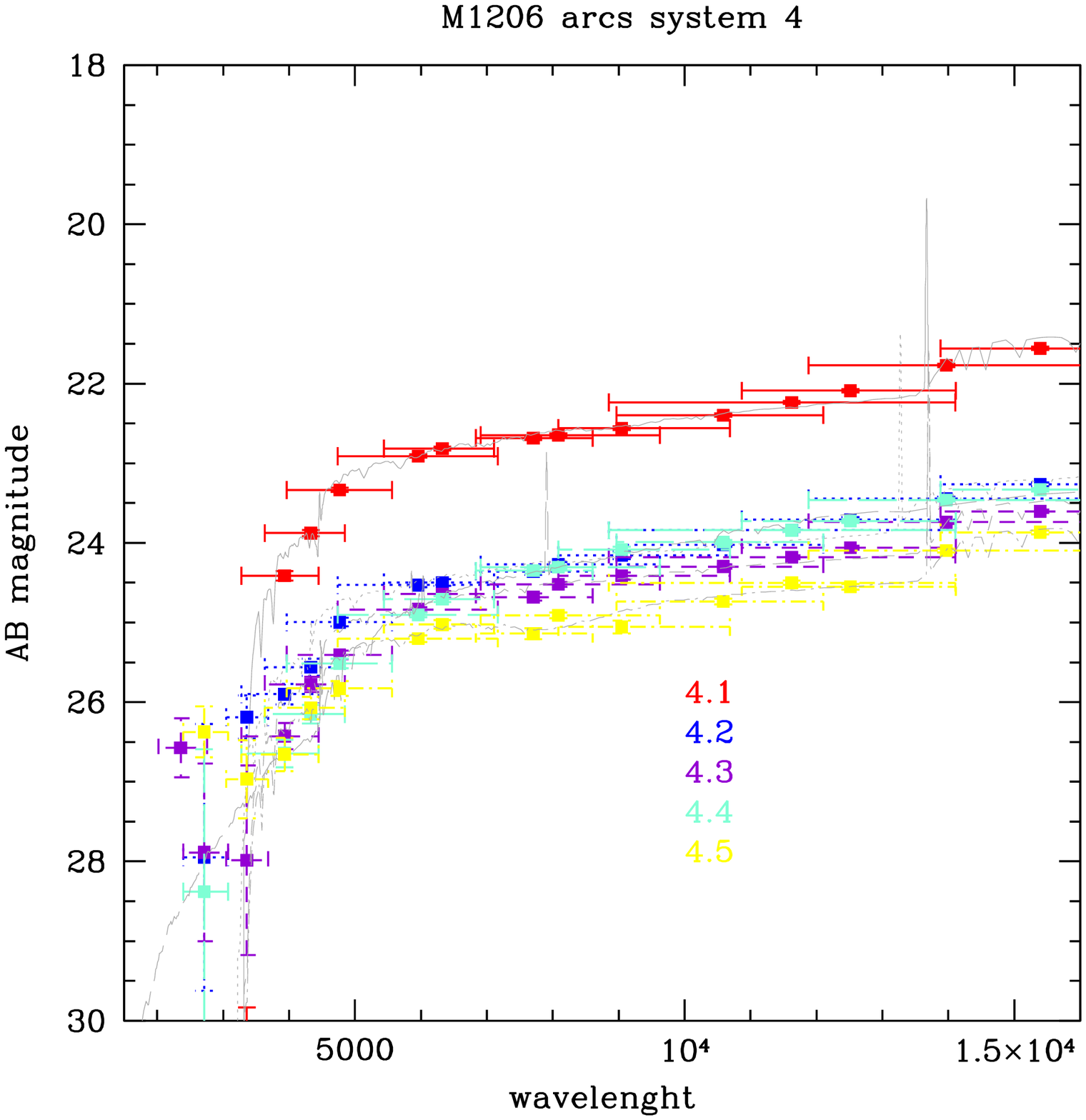}
\includegraphics[width=0.35\textwidth,height=0.3\textwidth]{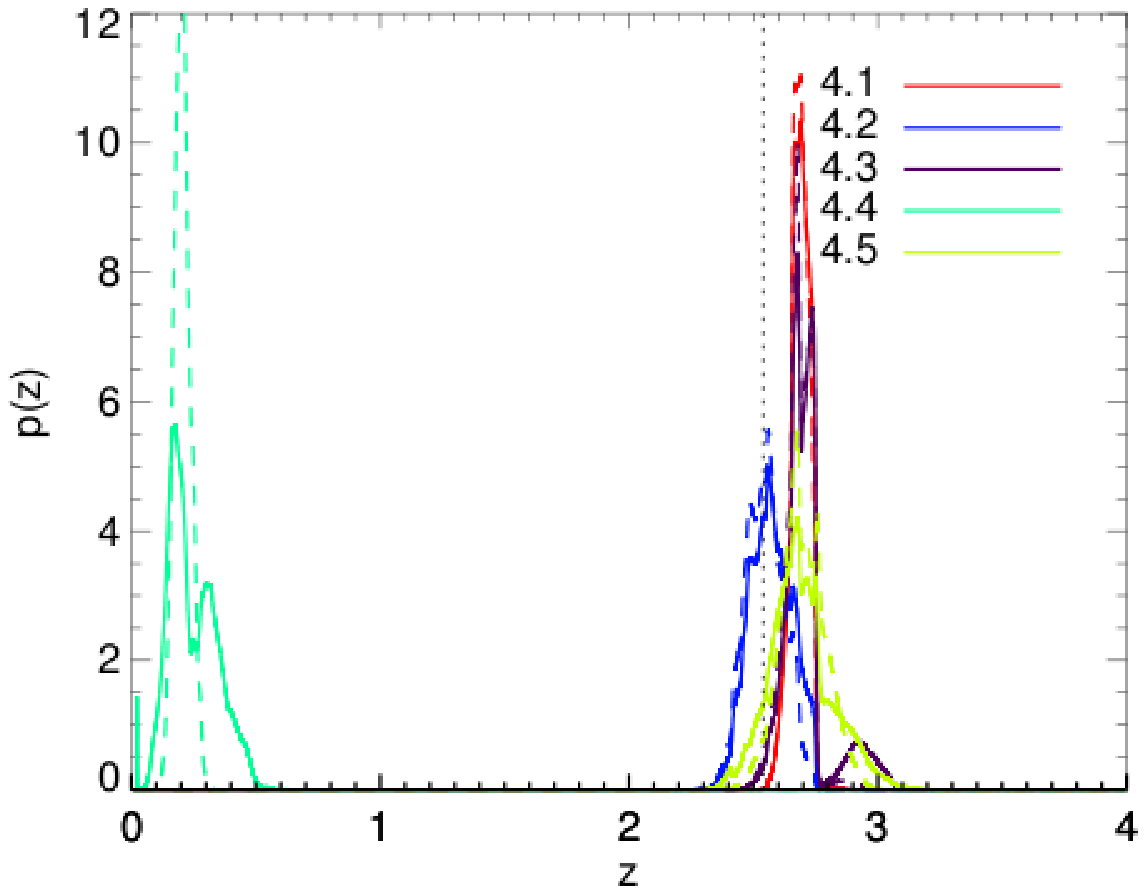}}
\caption{Overview of the multiply imaged sources with spec-z in MACS1206. 
The left panels show photo-z best-fit templates of each strong-lensing image from Le Phare, and the CLASH/HST photometry.
The vertical errorbars are photometric errors and the horizontal are the width of the HST filters.
In the right panels, the solid lines are the marginalized Le Phare $p(z)$
and the dashed lines are the $p(z)$ obtained by only considering the best-fit template in each case.
We assume flat priors in redshift and template space.
}
\label{fig:spectra_pz_sys}
\end{figure*}

\autoref{fig:spectra_pz_sys} shows, in the top row, the HST-image cutouts of some of the multiple image systems 
with the aperture that we used to derive the photometry of the
different arcs. We note that each HST-image cutout shows several arcs from differents systems 
while the middle and bottom row show a different arcs system at each column. 
\noindent

For arc 1.1, custom photometry gives more accurate photo-z results
than the standard isophotal photometry with the difference between the spec-z and 
the best-fit photometric redshift at less than 3\%(1+z) for both BPZ and Le Phare.
For arc 1.2, the custom aperture and the SExtractor-generated isophotal aperture 
are similar and the photo-z for both apertures agrees with the spec-z value to well within 3\%(1+z).
For arc 1.3, the isophotal photometry gives more accurate photo-z results than the custom photometry for both
Le Phare and BPZ. The isophotal photometry draws a smaller aperture around arc 1.3
that yields higher precision colours which improves the photo-z results.

System 1 shows good photo-z results from both BPZ and Le Phare
at less than 3\%(1+z) with higher accuracy when using smaller
aperture photometry, custom photometry for arc 1.1 and isophotal photometry 
for arc 1.3.
All best-fits templates are starbursts galaxies. 
The lensed background galaxies are usually very blue young galaxies and are thus best-fitted by starburst templates.
One would however need an individual template fitting procedure
to determine the type of each of these galaxies.

\begin{table}[!ht]
\caption{Photo-z for the multiple images with spec-z in MACS1206, for the custom tailored photometry and the standard isophotal photometry.}
\begin{tabular}{ccccccc} \hline\hline
 & arc & LP $z_{best}$ - $pdz\_best$ & BPZ $z_{best}$ - $odds$ & F814W\\
\hline\hline
custom & 1.1 & 1.04 $_{ -0.03 }^{ 0.07 }$ - 100 & 0.96 $_{ -0.01 }^{ 0.02 }$ - 98 & 20.8\\
$z_s=1.033$   & 1.2 & 1.01 $_{ -0.04 }^{ 0.05 }$ -  100 & 0.98 $_{ -0.02 }^{ 0.01 }$ - 99 & 20.4\\
   & 1.3 & 0.83 $_{ -0.01 }^{ 0.24 }$ -   53 & 0.97 $_{ -0.07 }^{ 0.02 }$ - 95& 20.6\\

\hline
isophotal &1.1  &   0.85 $_{ -0.01 }^{ -0.01 }$ - 99 & 0.74 $_{ -0.07 }^{0.07 }$ - 100 & 20.6\\
 & 1.2  &   1.07 $_{ -0.03 }^{ 0.02 }$ - 100 & 1.04 $_{ -0.08 }^{0.08 }$ - 100 &  21.0 \\
 & 1.3  &   1.05 $_{ -0.02 }^{ 0.03 }$ - 100 & 1.05 $_{ -0.08 }^{0.08 }$ - 100 & 22.4 \\

\hline\hline
custom & 2.1 & 3.18 $_{ -0.03 }^{ 0.04 }$ - 100 & 3.367 $_{ -0.004 }^{ 0.011 }$ - 100 & 23.5 \\
$z_s=3.03$ & 2.2 & 3.25 $_{ -0.10 }^{ 0.02 }$ - 100 & 3.39 $_{ -0.02 }^{ 0.01 }$ - 100 & 23.9\\
  & 2.3 & 3.22 $_{ -0.05 }^{ 0.06 }$ - 100 & 3.387 $_{ -0.022 }^{ 0.002 }$ - 100 & 24.1 \\
\hline
isophotal & 2.1  &  3.21 $_{ -0.07 }^{ 0.05 }$ - 99  & 0.41 $_{ -0.06}^{0.06}$  - 100 & 22.9\\
 & 2.2  &   3.20 $_{ -0.09}^{ 0.20 }$  -  98 & 0.41 $_{ -0.06}^{0.06}$ - 100  & 22.9\\
 & 2.3  &   3.64 $_{ -0.06 }^{ 0.08 }$  - 99 & 3.68 $_{-0.18}^{ 0.18}$ - 100 & 23.6\\
\hline\hline
custom & 3.1 & 3.21 $_{ -0.05 }^{ 0.06 }$ - 100 & 3.367 $_{ -0.044 }^{ 0.003 }$ - 100 & 24.0\\
$z_s=3.03$ & 3.2 & 3.17 $_{ -0.06 }^{ 0.07 }$ - 100 & 3.25 $_{ -0.02 }^{ 0.03 }$ - 99 & 24.0\\
 & 3.3 & 3.26 $_{ -0.08 }^{ 0.14 }$ -  100 & 3.39 $_{ -0.03 }^{ 0.02 }$ - 100 & 24.5\\
\hline
isophotal & 3.1  &   3.65 $_{ -0.06 }^{ 0.09 }$ - 99  & 3.73 $_{-0.19}^{  0.19 }$ - 100 & 23.9\\
 & 3.2  &   2.62 $_{ -0.13}^{ 0.47 }$ - 94 & 0.13 $_{-0.08}^{  3.09 }$ -  83 & 23.0\\
 & 3.3  &   3.62 $_{ -0.12 }^{ 0.06 }$ - 95 & 3.52 $_{-0.18}^{  0.18 }$ - 98 &  23.9\\
\hline\hline
custom & 4.1 & 2.67 $_{ -0.08 }^{ 0.08 }$ - 100 & 3.05 $_{ -0.03}^{ 0.02 }$ - 100 &  22.6\\
$z_s=2.54$ & 4.2 & 2.56 $_{ -0.15 }^{ 0.19 }$ - 100 & 3.09 $_{ -0.04 }^{ 0.03 }$ - 96 & 24.2\\
 & 4.3 & 2.68 $_{ -0.09 }^{ 0.08 }$ - 100 & 3.19 $_{ -0.06 }^{ 0.03 }$ - 93 & 24.4\\
 & 4.4 & 0.21 $_{ -0.12 }^{ 0.18 }$ -  78 & 0.21 $_{ -0.02 }^{ 0.06 }$ - 49 & 24.1\\
 & 4.5 & 2.68 $_{ -0.27 }^{ 0.24 }$ -  99  & 3.03 $_{ -0.12 }^{ 0.04 }$ - 73 & 25.1\\
\hline
isophotal & 4.1  &   2.54 $_{ -0.08 }^{ 0.07 }$ - 100  & 2.99 $_{-0.16}^{  0.16}$ - 100 &22.2\\
 & 4.2  &   2.35 $_{ -0.19 }^{ 0.33 }$ - 72  & 2.54 $_{-0.18}^{  0.17}$ - 91 & 24.1\\
 & 4.3  &   1.93 $_{ -0.02 }^{ 0.29 }$ - 61 & 2.35 $_{-0.19}^{  0.19}$ - 85 & 24.3\\
 & 4.4  &   0.520 $_{ 0.000 }^{ 0.005 }$ - 145 & 0.36 $_{-0.05}^{  0.07}$ - 97 & 20.6\\
 & 4.5  &   2.67 $_{ -0.21 }^{ 0.60 }$ - 65 & 3.04 $_{-0.21}^{  0.16}$ - 92 &  24.7\\
\hline\hline

\end{tabular}
\label{tab:pz_sys}
\end{table}

System 2 has three arcs from a source galaxy at redshift 3.03 for which spectroscopy
has been obtained from the on going VLT campaign.
We have good photometry in the visible and NIR for all the arcs.

The photo-z of the three images of system 2 have 2$\sigma$ confidence region that
are off by 0.15 in redshift.
The Balmer break being out of the
CLASH filter range, the Lyman break is the only strong colour
that will help the fitting as shown in \autoref{fig:spectra_pz_sys}.
The custom photometry shows better photo-z results than the isophotal for all arcs
with more consistent photo-z values closer to the spectroscopic
redshift even if off by at least 0.15 in redshift.

As for system 2, we do not find a lot of flux in the UVIS filters. 
The photometric redshift of system 3 is very similar to system 2 and
the images are radially very close to each other. Systems 2 and 3 are probably a galaxy group lensed
by MACS1206.

Arcs of system 3 have very similar photometry except in the bluest filter. They show
similar photo-z fit as you can see in \autoref{fig:spectra_pz_sys} and Table \ref{tab:pz_sys}.
The custom photometry gives photo-z estimates closer to the spec-z than the isophotal photometry even if 
still off by at least 0.2 in redshift.  
The standard photometry shows photo-z results that are off by 0.5 in redshift and values that
are not always consistent for all the arcs. BPZ and Le Phare show similar results for both
photometry.

Arcs of system 4, as for systems 2 and 3, do not have much flux in the UVIS
filters. However, the Balmer break lies in the NIR filters and improves the reliability of 
the high redshift solution for arcs 4.1, 4.2, 4.3 and 4.5. 
Arc 4.4 is very close to the MACS1206 BCG and is likely contaminated by the BCG light as
illustrated in the bottom panel right figure of \autoref{fig:spectra_pz_sys}.
This issue will be adressed in a further paper \citet{Molino13}.
The photo-z of arc 4.1, 4.2 and 4.5 are at less than 2$\sigma$ from
the spectroscopic redshift and at less than 3\%(1+z). All images
except 4.4 have good fits and are very close to the
spectrocopic redshift, for Le Phare using the custom photometry. BPZ shows slightly
biased high photometric redshift for the custom photometry.

For all the systems, BPZ and Le Phare show very close results for most cases
and at less than 2\%(1+z). Both BPZ and Le Phare give consistent and better
results using the custom photometry compared to the standard isophotal photometry
since we usually draw smaller tailored apertures compared to the standard one. Smaller aperture photometry
shows higher colour accuracy and better photo-z from Le Phare and BPZ for all arcs.

\section{MACS1206 photometric redshift catalogue}
\label{sec:photo-z_cat}

We release a photo-z catalogue of all sources in MACS1206 CLASH/HST field.  
This catalogue includes the multi-band photometry information (including upper-limits) 
and BPZ photo-z for each object that
has been detected as described in section \ref{sec:clash}. This catalogue is currently online
in the Mikulski Archive for Space Telescopes (MAST) system \footnote{http://archive.stsci.edu/prepds/clash/}. 
To the online MAST catalogue, we add our new Le Phare photo-z. 

There are a total of 3510 objects in the catalogue, in an area of
4.08 arcmin$^2$ covered by the CLASH programme.
We give the coordinates of each objects along with other SExtractor output
such as an estimation of the size given by the full-width-half-maximum, the area 
used to derive the photometry, an estimate of the ellipticity calculated 
from the second moment of the light distribution \citep{Bertin96} 
and a probability of the object being a star called stellarity, determined by 
a neural network approach.
For the photo-z results, we give the best-fit redshift
with a 68 and 95\% confidence region for both BPZ and
Le Phare. We also include the $odds-pdz\_best$ values in each case.
\begin{figure}[!ht]
\resizebox{\hsize}{!}{\includegraphics{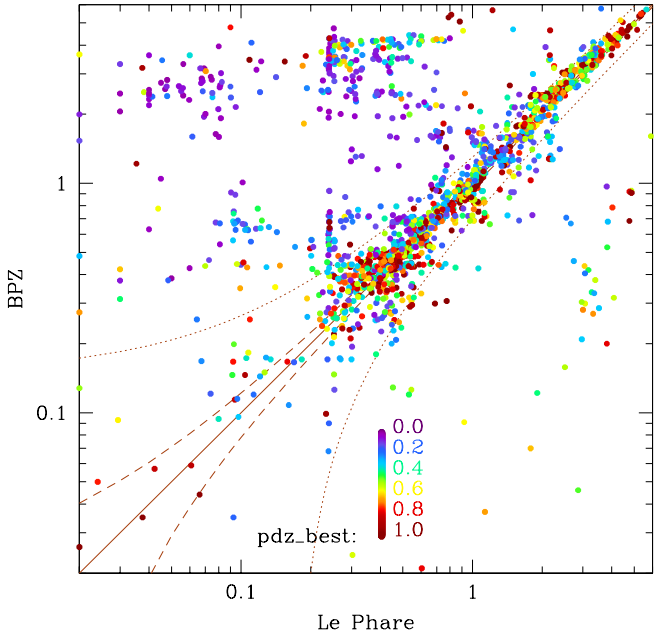}}
\caption{Comparison between Le Phare and BPZ photo-z in MACS1206 field
for galaxies with $S/N>10$ in F814W and stellarity lower than 0.08. Colors scale with the $pdz\_best$ parameter, taken from Le Phare.
The dashed and dotted lines represent respectively 2\%(1+z) and 15\%(1+z). Le Phare
and BPZ agrees well except a small fraction of faint galaxies with low confidence
photo-z that Le Phare places at low redshift and BPZ spreads over the redshift
range.}
\label{fig:bpzlpz_m1206}
\end{figure}

\autoref{fig:bpzlpz_m1206} shows Le Phare and BPZ photo-z in the MACS1206
field for galaxies with $S/N>10$ in F814W and a stellarity lower than 0.08. Stellarity
is close to 0 for galaxies and goes up to 1 for stars. In MACS1206 case, a stellarity lower than 0.08
select 86\% of the 10$\sigma$ F775W detected sample. 
Le Phare and BPZ agree very well. 
There are discrepancies for some galaxies
placed at low redshift by Le Phare but at high redshift for BPZ. These
discrepancies represent 7\% of the galaxies with $S/N>10$ and happen
where we usually find catastrophic outliers due to the confusion
between the Lyman and Balmer breaks.  
Those galaxies will likely yield poor photometric redshifts although no definitive conclusion can
be drawn without the spec-z information.
Most of the galaxies for which BPZ and
Le Phare disagree have $pdz\_best$ lower than 0.4, and so they can be pruned
with an $pdz\_best$ cut.

\begin{figure}[!ht]
\resizebox{\hsize}{!}{\includegraphics{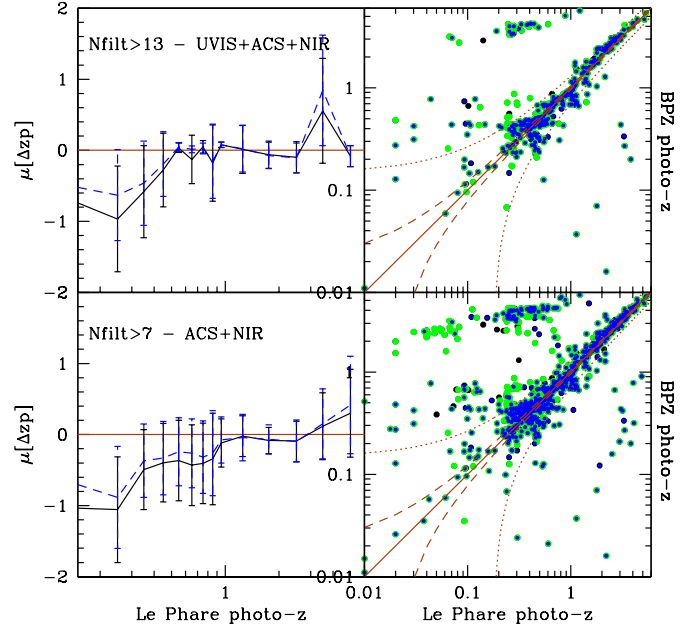}}
\caption{Comparison between Le Phare and BPZ photo-z in MACS1206 field
for galaxies with $S/N>10$ in F814W and stellarity lower than 0.08. 
Figures on the left show the mean value of the $\Delta zp=z_p^{Le Phare}-z_p^{BPZ}$
as a function of $z_p^{Le Phare}$ for ACS+NIR and UVIS+ACS+NIR photometry
respectively bottom and top panels. The blue and black lines are respectively
Le Phare and BPZ photo-z. The error bars are the rms between Le Phare and BPZ photo-z.
Right panels show the Le Phare vs BPZ photo-z. The green and blue dots show
a selection of respectively BPZ $\chi^2$ and Le Phare $pdz\_best$ 85\% highest
confidence redshifts. 
The dashed and dotted lines represent respectively 2\%(1+z) and 15\%(1+z). }
\label{fig:bpzlpz_nf_zp_m1206}
\end{figure}
\autoref{fig:bpzlpz_nf_zp_m1206} shows different selections of galaxies
to have some understanding of the photo-z quality in comparing Le Phare
and BPZ. The sample UVIS+ACS+NIR and ACS+NIR is made from galaxies with
at least respectively 13 and 7 filters detection. From the number of filter
selection, we conclude that galaxies with a high number of detections are
likely to have a good photo-z estimates. We also do a pre-selection of the highest
confidence 85\% galaxies from BPZ $\chi^2$ and Le Phare $pdz\_best$ and
eliminate most of the BPZ-Le Phare outliers. The corresonding values of 
($\chi^2$, $pdz\_best$) are (1.44,37). For more details about the performance
of these selections, see the Appendix \autoref{fig:bpzlpz_nf_mag_m1206} which
shows the selections as a function of magnitude and their completness as a function
or magnitude and redshift. 

\begin{figure}[!ht]
\resizebox{\hsize}{!}{\includegraphics{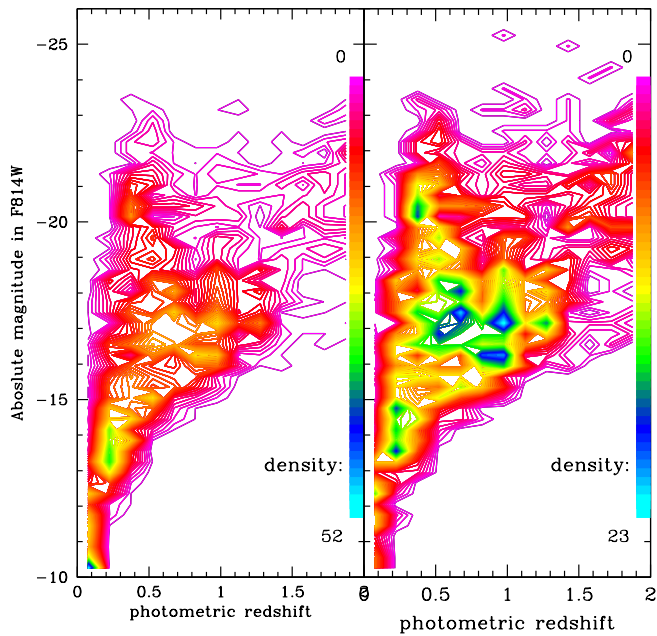}}
\caption{Contour showing the galaxy number density as a function of the absolute magnitude
in F814W and redshift for MACS1206 galaxy cluster. The absolute magnitude is 
derived from Le Phare photo-z fitting. The left and right panels have respectively
no $pdz\_best$ cut and a cut at 0.1. The galaxy number density is in linear colour scale.
The low redshift and intrisically faint galaxies show a poor photo-z estimation with
$pdz\_best<$0.1.} 
\label{fig:mabsz}
\end{figure}

\autoref{fig:mabsz} shows the galaxy number density for MACS1206 galaxy cluster.
The left panel of \autoref{fig:mabsz} shows a high density of faint galaxies
at low redshift.  On the right panel of \autoref{fig:mabsz}, we show the
same absolute magnitude-redshift-number density relation with a photoz quality cut
of $pdz\_best>0.1$. The quality cut removes most of the faint low redshift galaxies.
The faint low redshift galaxies are just an artefact 
from a bad fitting due to faint poorly constrained photometry.
 
\begin{figure}[!ht]
\resizebox{\hsize}{!}{\includegraphics{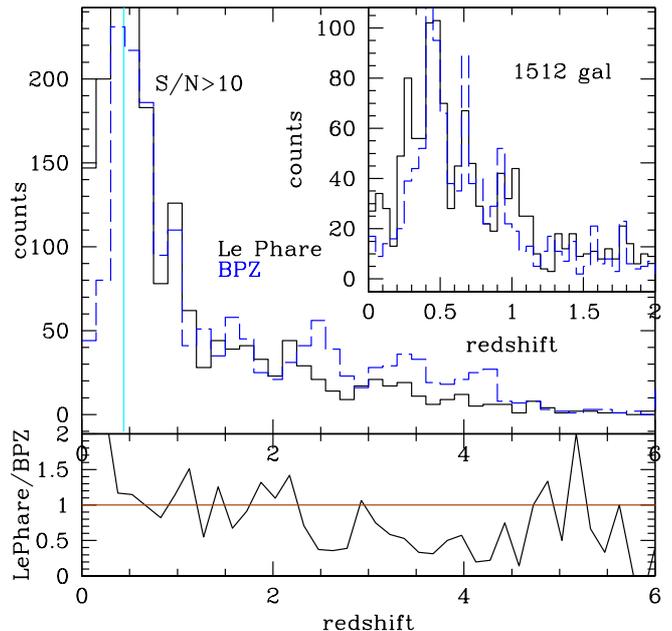}}
\caption{Photometric redshift histogram of MACS1206 using BPZ in blue dashed
line and Le Phare in black solid line. The cyan solid line is MACS1206 redshift at
$z\sim0.44$. BPZ and Le Phare have similar redshift distribution and show a well defined peak
at the cluster location. The top right corner panel displays BPZ and Le Phare photo-z
histograms for a selection of $odds$ and $pdz\_best$ values in order to select the best
85\% of galaxies for both BPZ and Le Phare.}
\label{fig:m1206_cluster}
\end{figure}

\autoref{fig:m1206_cluster} shows photo-z histograms of galaxies with 
a $S/N>10$ in F814W of Le Phare in black solid line and BPZ in blue dashed line. 
The cyan solid curve is at MACS1206 cluster redshift.
We remove the stars using SExtractor stellarity flag. 
The main figure shows BPZ and Le Phare photo-z with the signal-to-noise and stellarity selection. 
The top right corner histograms show BPZ and Le Phare photo-z with a quality cut 
which selects 85\% of the highest confidence 
photo-z of the galaxy sample $(pdz\_best,odds>0.15,0.31)$ for the top right corner histograms.
The bottom panel of \autoref{fig:m1206_cluster}
shows ratio of Le Phare and BPZ histograms. BPZ and Le phare agree well
from $z_p>0.2$ up to redshift $z_p<2$. At $z_p>2$ BPZ finds more galaxies than Le Phare. 
As described in the above paragraph, we note that Le Phare has an unrealistic 
number of galaxies at low redshift $z_p<0.2$ that are placed at high redshift
by BPZ that can be pruned with an odds cut.  

\begin{figure}[!ht]
\resizebox{\hsize}{!}{\includegraphics{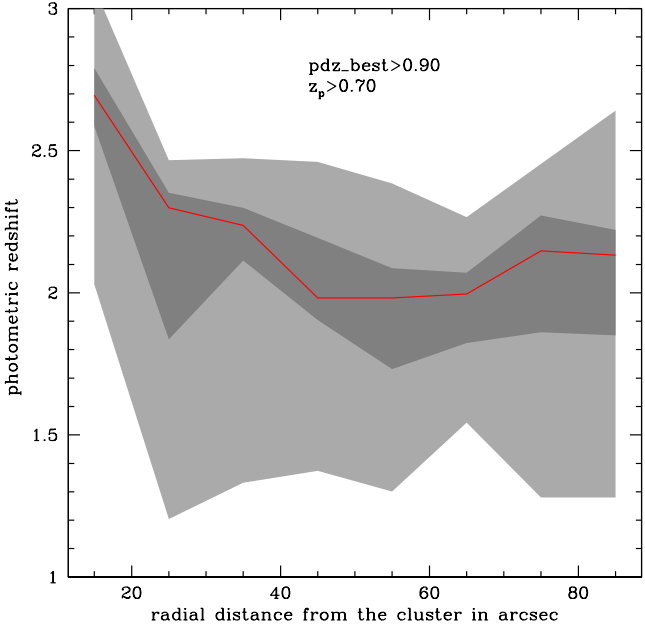}}
\caption{Mean photo-z of galaxies located behind MACS1206 
in solid red line as a function of angular radial distance from the cluster centre. 
The dark grey area reprensent the mean 68\% area around the mean and the light grey the 95\% area.
Photo-z are derived from Le Phare code.} 
\label{fig:zmean}
\end{figure}

\autoref{fig:zmean} shows the mean photo-z as a function of cluster-centric radius 
of galaxies located behind MACS1206 with a S/N$>$10 in F775W, stellarity lower than 0.08. 
It shows a mean redshift higher close to the centre of the cluster due to
the greater magnification near the cluster centre. In the radial bin closest to the cluster
centre, the mean redshift is around 2.5, while for the radial bins further away the
mean redshift is around 2 and does not change much 
as we go further away from the cluster centre. 

\section{Sensibility of the mass reconstruction from SL to the photo-z accuracy}
\label{sec:mass_rec}

\begin{figure}
\begin{center}
\includegraphics[width=\columnwidth]{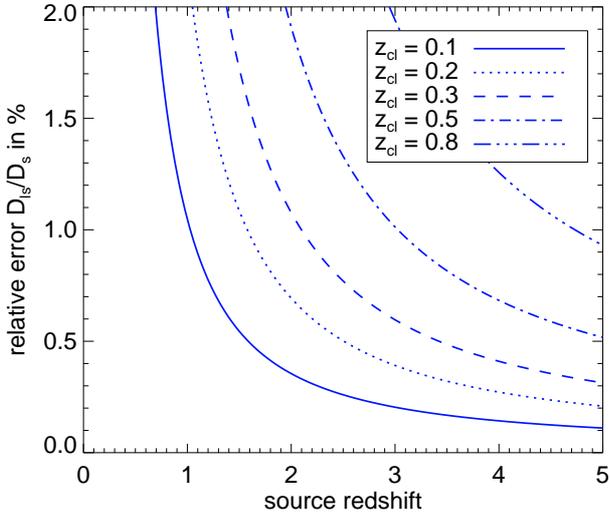}
\caption{The relative error on the distance ratio $D_{ls}/D_s$ as a function of source redshift $z_s$ for
various lens redshifts $z_{cl}$, assuming a photo-z precision of 4\%}
\label{fig:relerr}
\end{center}
\end{figure}

Now we consider the impact of the photo-z uncertainty on the lens modelling. The accuracy of the photo-zs measured for
arcs and lensed images directly influences the accuracy of the mass model through the lens equation,
which connects the unknown source position $\beta$ with the image position $\theta$
\begin{equation}
{\beta}={\theta}-\frac{D_{ls}}{D_s}{\hat{\alpha}}({\theta}).
\end{equation}
The mass model determines the deflection field ${\hat{\alpha}}$ and the photo-z enters in the
distance ratio $D_{ls}/D_s$. \autoref{fig:relerr} shows how the relative photo-z accuracy is mapped
to the distance ratio using standard linear error propagation. It is clear that low redshift sources are
the most challenging. The relative accuracy of the distance ratio can be regarded as the relative error on
the overall mass scale of the lens since $\hat{\alpha}\propto M$ so, e.g., a z=2 source lensed by a z=0.3
cluster will give rise to a relative uncertainty on $D_{ls}/D_s$ and on the total mass of $\sim$1\% if
the photo-z uncertainty is 4\%. However, the exact influence of the photo-z accuracy on the lens model will
depend on the number, positions, and redshifts of the lensed images in a complex way, and it may also
depend on whether the lens model is parametric or not. This will be assessed on a case--by--case basis.

\begin{figure}
\begin{center}
\includegraphics[width=\columnwidth]{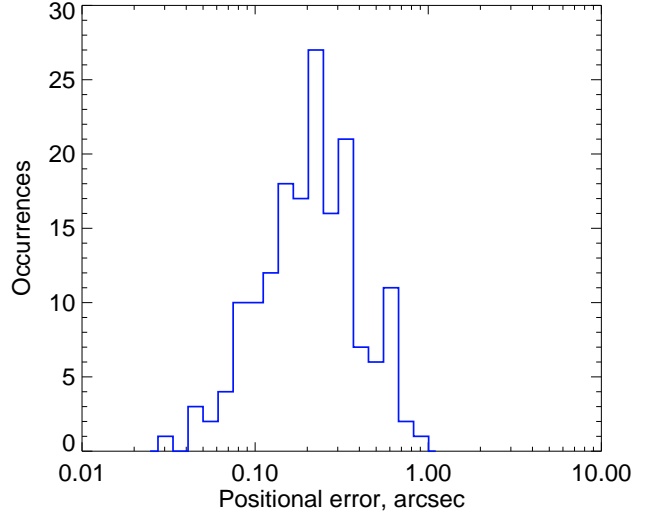}
\caption{The distribution of photo-z positional errors for the 168 images lensed by Abell 1689. }
\label{fig:a1689}
\end{center}
\end{figure}
However, we can also regard the photo-z uncertainty as a positional uncertainty in the image plane,
$\delta \alpha_z$, through the lens equation. The relative accuracy of the distance ratio in
\autoref{fig:relerr} is then equal to the relative accuracy on the distance between the image
and the model-predicted source position. This approach facilitates comparison with other positional
uncertainties that affect the lensing analysis. To be specific, we base our comparison on the 168
lensed images produced by Abell 1689 and their corresponding source positions as found using the
lens model of \citet{Coe10}. We imagine that each image has a photo-z
and that the overall photo-z precision is $4\%$. The disttribution of the photo-z positional uncertainties,
which we label $\delta \alpha_z$ are shown in \autoref{fig:a1689}. We take this distribution
as a reasonable sample estimate of the impact of photo-z accuracy.

For \emph{HST} imaging, the accuracy of the astrometric solution is similar to the pixel scale,
$0.05$ arcsec for the ACS/HST camera. It is clear from \autoref{fig:a1689} that a 4\% photo-z precision
implies that, for most images, $\delta \alpha_z$ will be larger than the astrometric error.
However, larger positional errors are predicted: \citet{Jullo10} estimated that
images would pick up additional deflections of more than 1 arcsec from lensing by the large-scale
structure along the line-of-sight, and it is clear from \autoref{fig:a1689} that we expect
almost all $\delta \alpha_z$ to be smaller than this. In a more detailed matter power spectrum
analysis, \citet{Host12} showed that the impact of large-scale structure increases with
redshift and distance on the sky to the cluster centre, which means that low-redshift arcs will be
relatively more affected by the photo-z accuracy. We calculate the deflection $\delta \alpha_\mathrm{LSS}$
using the prescription of \citet{Host12}, and we find that
$\delta \alpha_z>\delta \alpha_\mathrm{LSS}$ for 6\% of the Abell 1689 images if the
photo-z precision is 4\% while this happens for only one image if the precision is 2\%. Hence, we conclude
that the impact of the photo-z accuracy of the mass model is most likely subdominant in comparison
to the uncertainty associated with lensing by large-scale structure along the line of sight if
we can achieve a photo-z precision of 4\% or better.
We note that this model is valid when the lensed images do not sit on a strong-lensing critical line.
In this last case, the non-linearity of equations makes any predictions difficult to realize.

Finally, there is the possibilty of catastrophic errors arising from the Lyman-Balmer break degeneracy,
which can give rise to two solutions in the photo-z fitting. However, this is usually a benign case for
the purpose fitting the lens model. It is extremely unlikely that the low-$z$ solution is a viable
redshift for a lensed image, in fact, for \emph{CLASH} clusters, it may well be a lower redshift than
the cluster lens itself. Even if the low redshift solution is viable for an arc it will be difficult to
fit the observed image positions since the separation between the lensed images would not accommodate
the low predicted $D_{ls}/D_s$ -- as long as there are other lensed images constraining the mass model.
Hence, catastrophic outliers will mostly be an issue for the identification of lensed images in the
first place, and in that case there are other constraints such as morphology which can be used.

\section{Conclusion} 
\label{sec:conclusion}
In this study, we evaluated the photo-z accuracy using 16 bands HST going from UV to NIR 
for 17 out of the 25 CLASH clusters.
We compare two photo-z methods, BPZ and Le Phare for all the galaxies in 17 cluster fields and test
them against spectrocopic redshifts. 
We studied separately the lensed background galaxies with a sample of 63 galaxies 
from the cluster member and foreground galaxies with 327 galaxies.
Our main conclusions are:\\
$\bullet$ Le Phare photo-z for the lensed background galaxies are somewhat 
more accurate than those for the foreground and cluster members when adding zeropoint adjustements. \\
$\bullet$ Using Le Phare, we reach a precision of 3\%(1+z) for the lensed background galaxies which is reduced to 2.4 \%(1+z)
after removing outliers (based on the quality of the fit). \\
Both BPZ and Le Phare have similar photo-z estimation for galaxies in the foreground and cluster field with a 
correponding values of 4.0 \%(1+z) and 3\%(1+z). 
This precision for the arcs satisfies the requirement for the mass model reconstruction as shown in section
\ref{sec:mass_rec}. Using a simple error propagation going from the photo-z to the lens model, we find a 1\% error 
on the derived lens model using a photo-z scatter of 4\%(1+z) for a cluster at z=0.3 and a source 
at z=2. This is however a simplistic model and a case by case study for each cluster is necessary for more
detail.\\
$\bullet$ The p(z)-based parameters $odds$ and $pdz\_best$ are demonstrated to be useful 
estimators of the quality of the derived photometric redshifts.\\ 
$\bullet$ Isophotal photometry, in general, performs better than other types of photometry 
although similar results are achieved with 20 pixels aperture photometry. \\
$\bullet$ We verified the above conclusions using mock galaxy catalogues that included varying degrees of complexity in the 
underlying galaxy physics.\\
$\bullet$ For MACS1206 we compare photo-z for the arcs using 2 photometric measurements and find that
using Le Phare and BPZ photo-z have a precision higher than 3\%(1+z). Smaller aperture photometry gives better photo-z results
usually obtained from the tailored aperture photometry. Photo-z results for the lensed background galaxies 
with the default SExtractor photometry show a scatter of 5\%(1+z) demonstrating the utility of the tailored photometry. 
For MACS1206 lensed systems 1, 2 and 4, Le Phare and BPZ agrees within 2$\sigma$ 
in redshift for the arcs and for the field galaxies 

We are also working on improving the photometry
with a background substraction model adjusted for each cluster using shapelets
\citep{Molino13}.

\begin{acknowledgements}
SJ and OH acknowledge STFC-supported Post-doctoral Felowships at UCL,
OL acknowledges a Royal Society Wolfson Research
Merit Award, a Leverhulme Senior Research Fellowship and
an Advanced ERC Grant. 
SJ is supported by the Spanish Science MinistryAYA2009-13936 Consolider- Ingenio CSD2007-00060, 
project2009SGR1398 from Generalitat de Catalunya and by the European Commissions 
Marie Curie Initial Training Network CosmoComp (PITN- GA-2009-238356).
The Dark Cosmology Centre is funded by the Danish National Research Foundation.
S. Seitz acknowledges support from the Transregional Collaborative Research Centre TRR 33 "The
Dark Universe" and from the DFG cluster of excellence
"Origin and Structure of the Universe".
This research is partially supported by PRIN INAF 2010: "Architecture and Tomography of Galaxy Clusters".
AF acknowledges the support by INAF through PRIN 2008 and 2010 grants.
Support for AZ is provided by NASA through Hubble Fellowship grant \#HST-HF-51334.01-A awarded by STScI. Part of this work was also supported by contract research ``Internationale Spitzenforschung II/2-6'' of the Baden W\"urttemberg Stiftung.
AM acknowledges support from AYA2006-14056BES-2007-16280.
\end{acknowledgements}

\bibliographystyle{aa}
\bibliography{biblio}
\appendix
\section{Spectroscopic redshifts and systematic shifts}

Table \ref{tab:cluster_spec-z} gives the number of very secure spectroscopic redshift 
available for galaxies that lie within the CLASH/HST fields of view.
Most spec-z's have been targeted by the VIMOS/VLT Large Program 186.A-0798 \citep{Rosati13}.
Other spec-z come from GISMO observations on Magellan telescopes, VLT observations \citep{Lamareille06},
the 6DF survey \citep{Jones04}, the SDSS DR7 \citep{Abazajian09}, the MMT/Hectospec survey \citep{Fabricant05,Rines13} 
and archival data from \citet{Ebeling09,Sand08,Smith05,Newman11,Richard11,Guzzo09,Halkola08,Cohen02,Stern10}. 
 
\begin{table}
\caption{Number of spectroscopic redshifts by cluster.}
\begin{tabular}{cccccccc} \hline\hline
Cluster & Nbr  \\
abell209  &     77\\
abell2261 &     16\\
abell383  &     63\\
abell611  &     33\\
macs0329  &      34\\
macs0416  &      8\\
macs0647  &      13\\
macs0744  &      8\\
macs1115  &      111\\
macs1206  &     147\\
macs1931  &      17\\
macs2129  &       36\\
rxj1347 & 76\\
rxj1532 & 3\\
rxj2129 & 28\\
\hline\hline
\end{tabular}
\label{tab:cluster_spec-z}
\end{table}

Using the spec-z listed in Table \ref{tab:cluster_spec-z}, we derive systematic shifts 
that we apply to the CLASH photometry as a first order template correction for Le Phare photo-z. 
These shifts are not applied to BPZ photo-z.
Using different sets of photometry and different types of photometry change the shift's values. 
In Table \ref{tab:shifts}, we list the shift's values from using the different sub-set
of the CLASH photometry that are defined in section \ref{subsec:zp_sys}.

\begin{table*}
\caption{Systematic shifts for sub-set of the CLASH filters as defined in section \ref{subsec:zp_sys}.}
\begin{tabular}{cccccccc} \hline\hline
Band& ACS & ACS+NIR & UVIS+ACS & UVIS+ACS+NIR & UVIS1+ACS+NIR & UVIS2+ACS+NIR & simul4\\
\hline\hline
F225W &   -  &   0.14 $\pm{ 0.46 }$ &     -  &   0.32 $\pm{ 0.57 }$ &     -  &    -  &   -0.11 $\pm{ 0.09 }$ \\
F275W &   -  &   0.28 $\pm{ 0.35 }$ &     -  &   0.43 $\pm{ 0.45 }$ &     -  &    -  &   -0.05 $\pm{ 0.07 }$ \\
F336W &   -  &   0.05 $\pm{ 0.23 }$ &     -  &   0.17 $\pm{ 0.26 }$ &     -  &   0.18 $\pm{ 0.27 }$ &    0.00 $\pm{ 0.06 }$ \\
F390W &   -  &   0.00 $\pm{ 0.11 }$ &     -  &   0.10 $\pm{ 0.12 }$ &    0.09 $\pm{ 0.12 }$ &    0.10 $\pm{ 0.12 }$ &    0.03 $\pm{ 0.05 }$ \\
F435W &  0.01 $\pm{ 0.06 }$ &    -0.03 $\pm{ 0.07 }$ &   0.05 $\pm{ 0.06 }$ &    0.03 $\pm{ 0.07 }$ &    0.03 $\pm{ 0.07 }$ &    0.04 $\pm{ 0.07 }$ &    0.03 $\pm{ 0.05 }$ \\
F555W &  0.03 $\pm{ 0.04 }$ &    -0.01 $\pm{ 0.04 }$ &   0.06 $\pm{ 0.04 }$ &    0.03 $\pm{ 0.05 }$ &    0.03 $\pm{ 0.04 }$ &    0.03 $\pm{ 0.04 }$ &    0.03 $\pm{ 0.04 }$ \\
F475W &  0.07 $\pm{ 0.04 }$ &    0.02 $\pm{ 0.04 }$ &    0.11 $\pm{ 0.04 }$ &    0.08 $\pm{ 0.06 }$ &    0.09 $\pm{ 0.05 }$ &    0.08 $\pm{ 0.05 }$ &    0.03 $\pm{ 0.03 }$ \\
F606W &  0.02 $\pm{ 0.03 }$ &    0.00 $\pm{ 0.03 }$ &    0.02 $\pm{ 0.03 }$ &    0.00 $\pm{ 0.03 }$ &    0.01 $\pm{ 0.03 }$ &    0.01 $\pm{ 0.03 }$ &    0.02 $\pm{ 0.04 }$ \\
F625W &  -0.01 $\pm{ 0.02 }$ &   -0.03 $\pm{ 0.02 }$ &   -0.01 $\pm{ 0.02 }$ &   -0.04 $\pm{ 0.02 }$ &   -0.02 $\pm{ 0.02 }$ &   -0.03 $\pm{ 0.02 }$ &   0.02 $\pm{ 0.04 }$ \\
F775W &  -0.03 $\pm{ 0.02 }$ &   -0.01 $\pm{ 0.02 }$ &   -0.05 $\pm{ 0.02 }$ &   -0.07 $\pm{ 0.02 }$ &   -0.05 $\pm{ 0.02 }$ &   -0.06 $\pm{ 0.02 }$ &   0.03 $\pm{ 0.04 }$ \\
F814W &  0.00 $\pm{ 0.01 }$ &    0.04 $\pm{ 0.02 }$ &    -0.02 $\pm{ 0.02 }$ &   -0.03 $\pm{ 0.02 }$ &   -0.02 $\pm{ 0.02 }$ &   -0.02 $\pm{ 0.02 }$ &   0.03 $\pm{ 0.04 }$ \\
F850LP&  -0.04 $\pm{ 0.02 }$ &   0.03 $\pm{ 0.02 }$ &    -0.07 $\pm{ 0.02 }$ &   -0.08 $\pm{ 0.02 }$ &   -0.06 $\pm{ 0.03 }$ &   -0.07 $\pm{ 0.02 }$ &   0.00 $\pm{ 0.04 }$ \\
F105W &  0.07 $\pm{ 0.03 }$ &     -  &   0.03 $\pm{ 0.02 }$ &    0.05 $\pm{ 0.02 }$ &    0.03 $\pm{ 0.02 }$ &    0.03 $\pm{ 0.02 }$ &    -0.02 $\pm{ 0.03 }$ \\
F110W &   -  &    -  &   0.00 $\pm{ 0.02 }$ &    0.03 $\pm{ 0.02 }$ &    -0.00 $\pm{ 0.02 }$ &   0.01 $\pm{ 0.02 }$ &    -0.00 $\pm{ 0.04 }$ \\
F125W &   -  &    -  &   0.06 $\pm{ 0.02 }$ &    0.09 $\pm{ 0.02 }$ &    0.06 $\pm{ 0.02 }$ &    0.07 $\pm{ 0.02 }$ &    -0.05 $\pm{ 0.04 }$ \\
F140W &   -  &    -  &   0.06 $\pm{ 0.03 }$ &    0.10 $\pm{ 0.03 }$ &    0.06 $\pm{ 0.03 }$ &    0.08 $\pm{ 0.03 }$ &    -0.07 $\pm{ 0.05 }$ \\
F160W &   -  &    -  &    -  &    -  &    -  &    -  &
\end{tabular}
\label{tab:shifts}
\end{table*}

\section{The $odds$ and $pdz\_best$ photo-z confidence values}

\begin{figure*}
\includegraphics[scale=0.45]{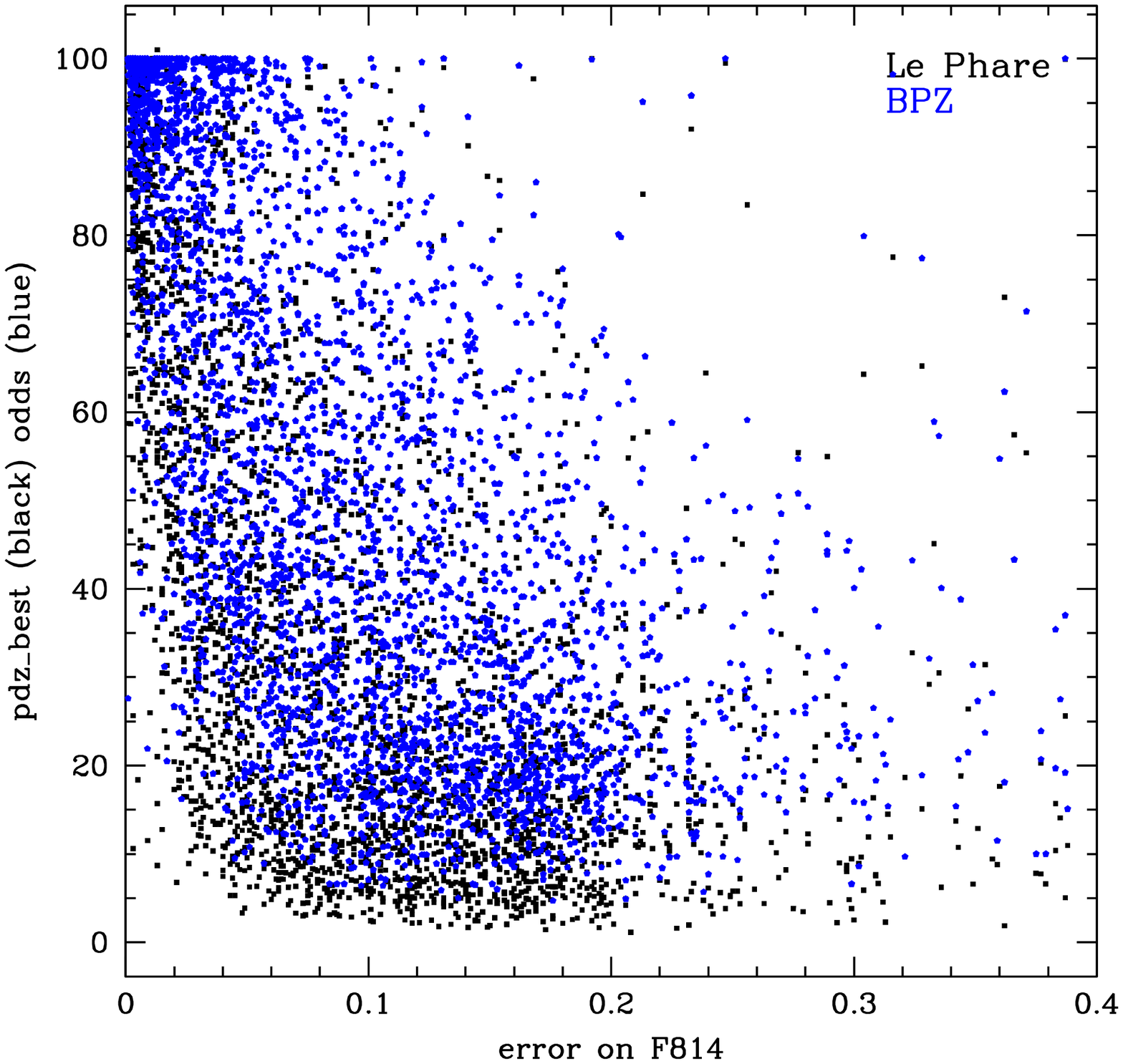}
\includegraphics[scale=0.45]{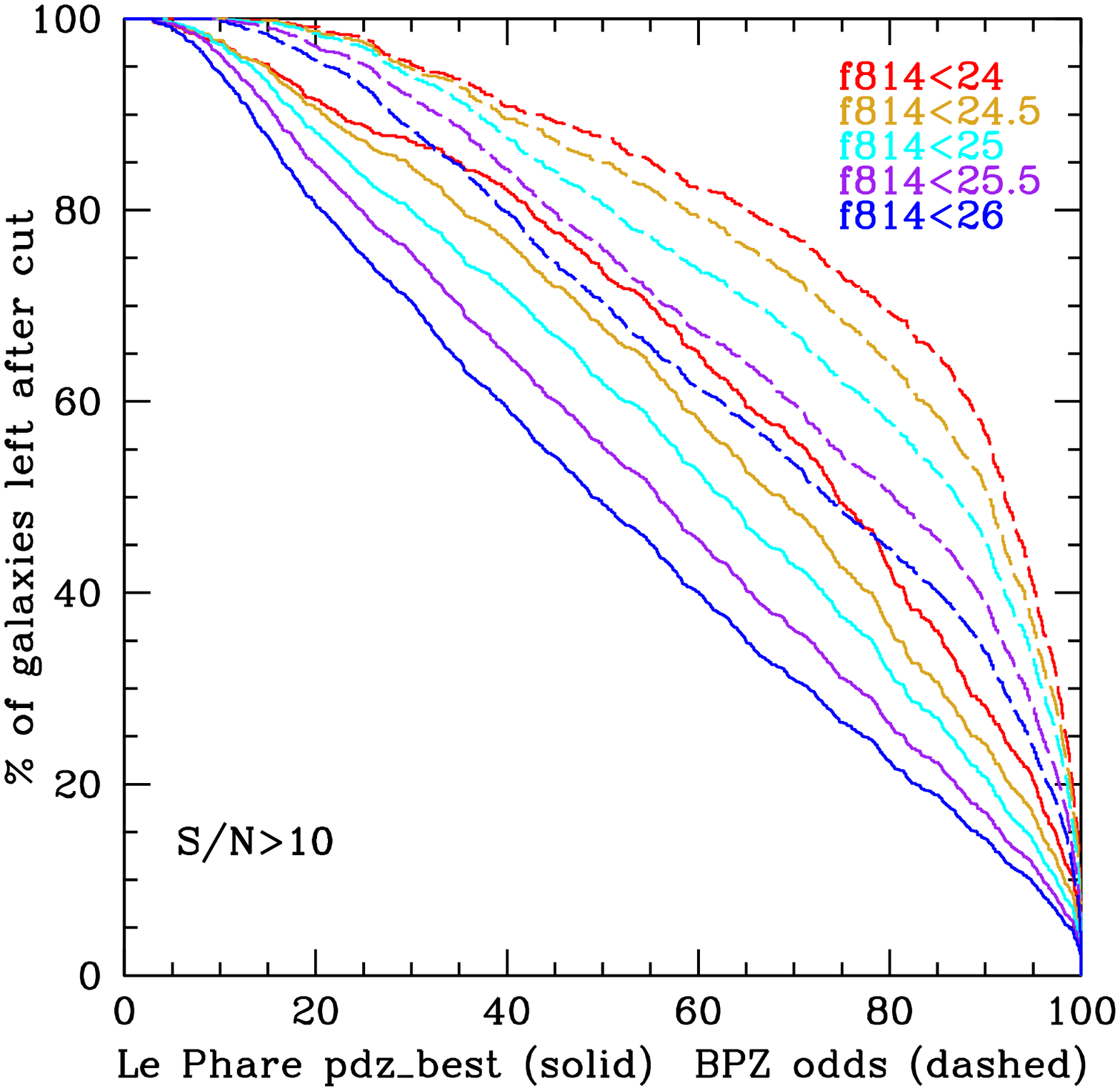}
\caption{Top figure shows the magnitude error as a function of $odds$ and $pdz\_best$ parameter for MACS1206 galaxy cluster. 
The black dots and blue stars show respectively Le Phare and BPZ results.
Bottom figure shows the percentage of galaxies left after a cut in the $pdz\_best$ parameter for Le Phare 
in solid lines and $odds$ parameter for BPZ in dashed lines for MACS1206 galaxy 
cluster at different F814W magnitudes limit.}
\label{fig:emag_odds}
\end{figure*}

The top panel of \autoref{fig:emag_odds} shows the error on the F814W magnitude
as a function of the $odds$ parameter of BPZ in black 
and $pdz\_best$ parameter of Le Phare in blue for MACS1206 galaxy cluster.
Bright galaxies have higher $odds$ than faint galaxies, as one can expect.   
We note that some Le Phare $pdz\_best$ values are higher than maximum possible
value 100. This is a numerical issue during the integration of the p(z). It
usually means that the p(z) is very narrow and can be assumed as high value
of $pdz\_best$.

The bottom panel of \autoref{fig:emag_odds} shows the percentage of galaxies left
after a selection cut based on $pdz\_best$ Le Phare in
solid lines and $odds$ BPZ in dashed lines at different F814W magnitude based
on MACS1206 galaxy cluster photoz for all galaxies in the HST field.
We note that to remove the same percentage of galaxies, one needs a higher
cut for the BPZ $odds$ parameter than with Le Phare $pdz\_best$.

\begin{figure}[!ht]
\resizebox{\hsize}{!}{\includegraphics{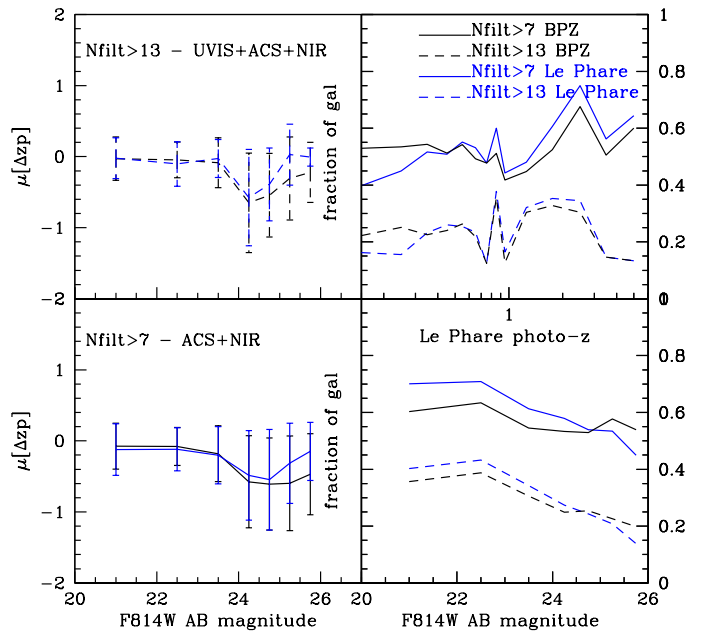}}
\caption{Le Phare and BPZ photo-z comparison for samples of galaxies
selected as a function of the number of detections using the best
85\% BPZ $\chi^2$ or Le Phare $pdz\_best$ values. The left panels show the
mean values of BPZ-Le Phare photo-z for the different samples. The right panels
show the fraction of galaxies selected by bins of magnitude and redshift.}
\label{fig:bpzlpz_nf_mag_m1206}
\end{figure}

\autoref{fig:bpzlpz_nf_mag_m1206} shows a comparison between Le Phare and BPZ
when selecting the best 85\% BPZ $\chi^2$ or Le Phare $pdz\_best$ as well as
a number of detections selection.

We note that BPZ $odds$ have values between 0 and 1 that we rescale at 0 to 100
in order to have the same scale between Le Phare $pdz\_best$ and BPZ $odds$.

\section{Arcs}
\label{app:arcs}

In this section, we show the photo-z results from the BPZ code for the standard isophotal photometry
for each of the lensed images of MACS1206 systems. These results are to be compared with those shown in
section \ref{sec:arcs} for Le Phare.
We show stamps images of each arcs, SExtractor segmentation map for the isophotal photometry,
the probability of the photo-z an the best-fit template of the BPZ results from the first left column 
to last column. The aperture shape of the SExtractor isophotal photometry is showed by the black contour
on the second column from the left. When comparing these figures to figure \autoref{fig:spectra_pz_sys}
we note that the tailored photometry and the isophotal SExtractor photometry have different shapes.
Photo-z derive from both shapes using the same photo-z code usually prefers the tailored photometry as
explained in section \ref{sec:arcs}.

\begin{figure*}[!ht]
\hbox{
\includegraphics[scale=0.85]{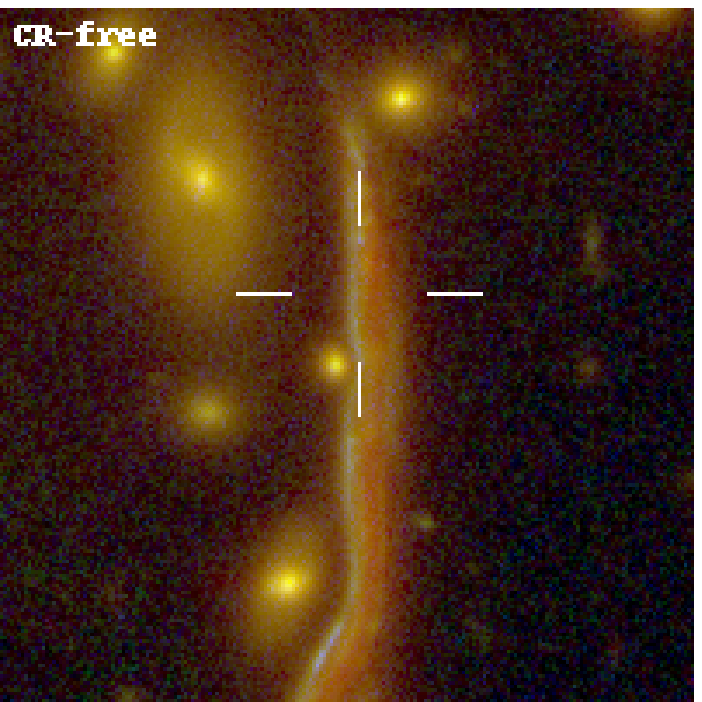}
\includegraphics[scale=0.85]{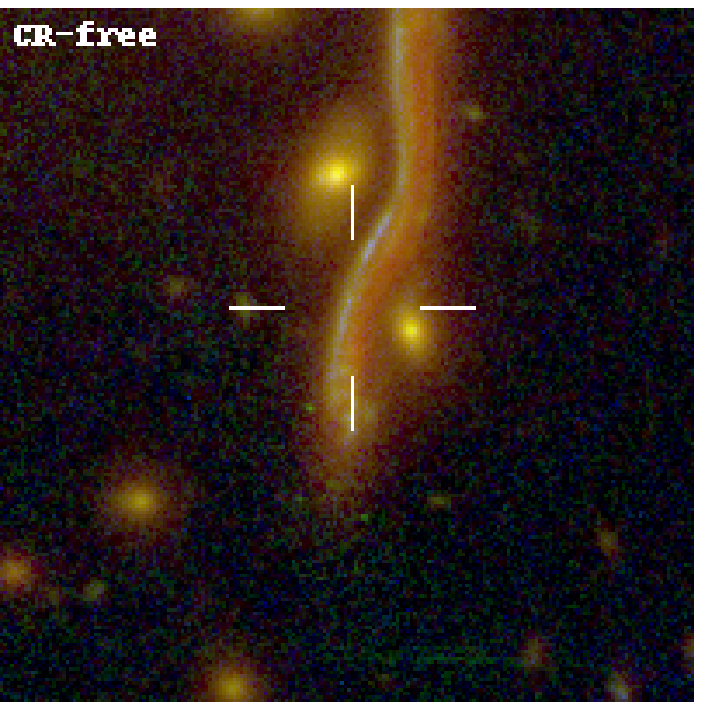}
\includegraphics[scale=0.85]{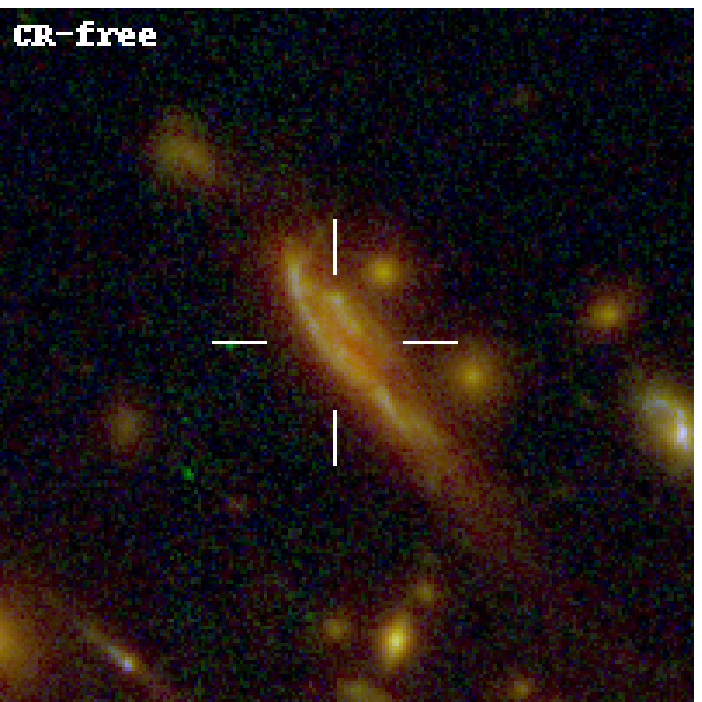}}
\hbox{
\includegraphics[scale=0.85]{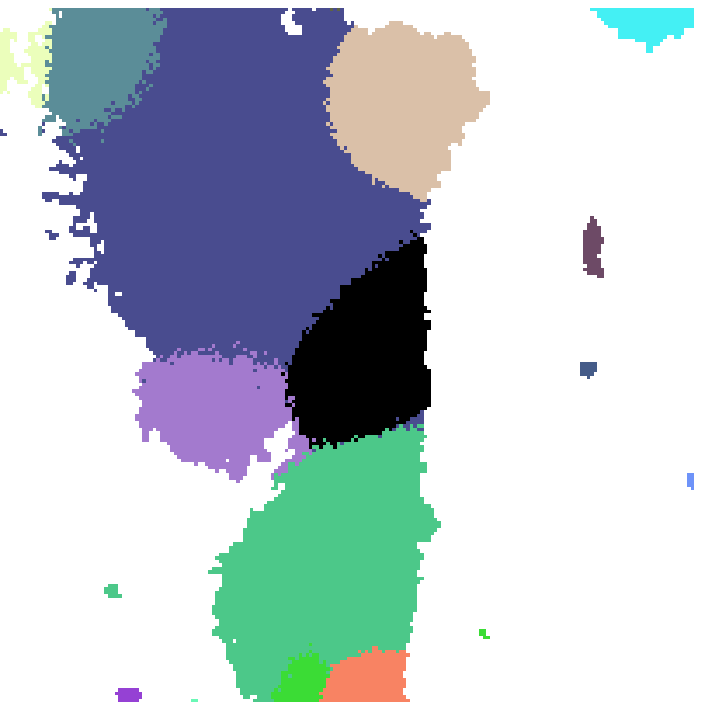}
\includegraphics[scale=0.85]{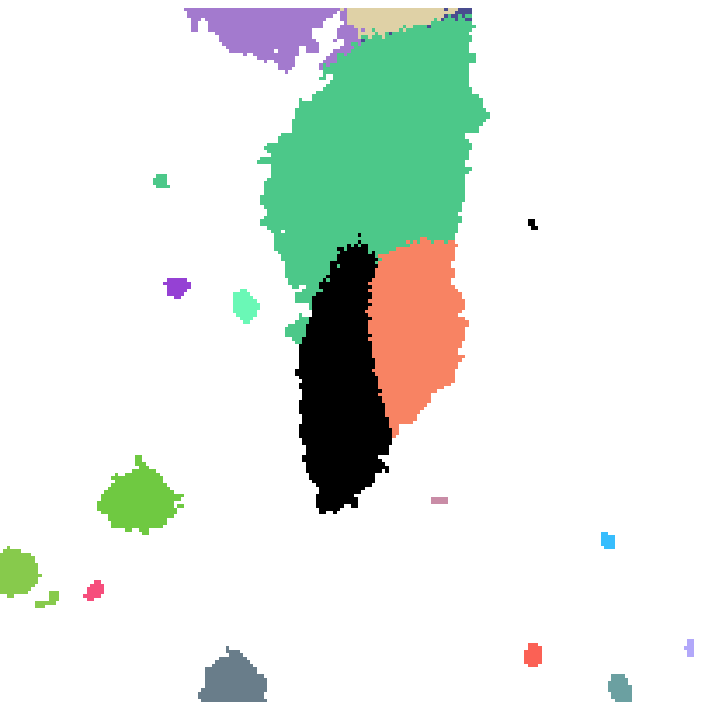}
\includegraphics[scale=0.85]{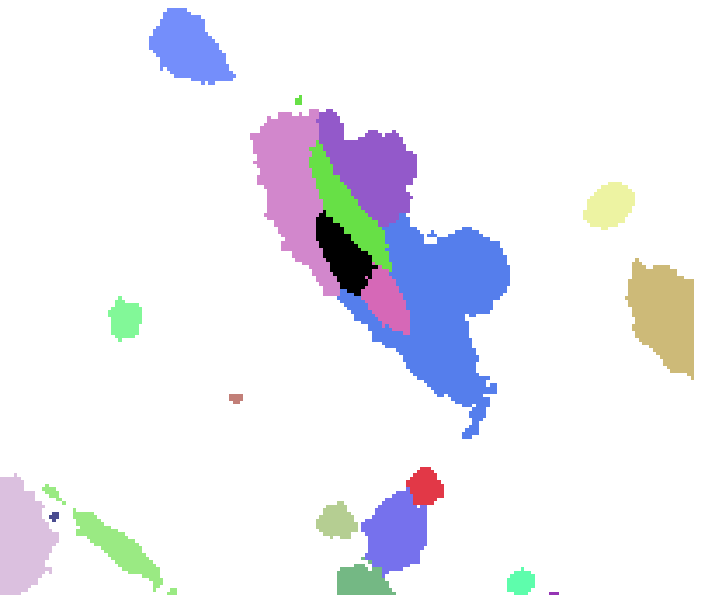}}
\hbox{
\includegraphics[scale=0.3]{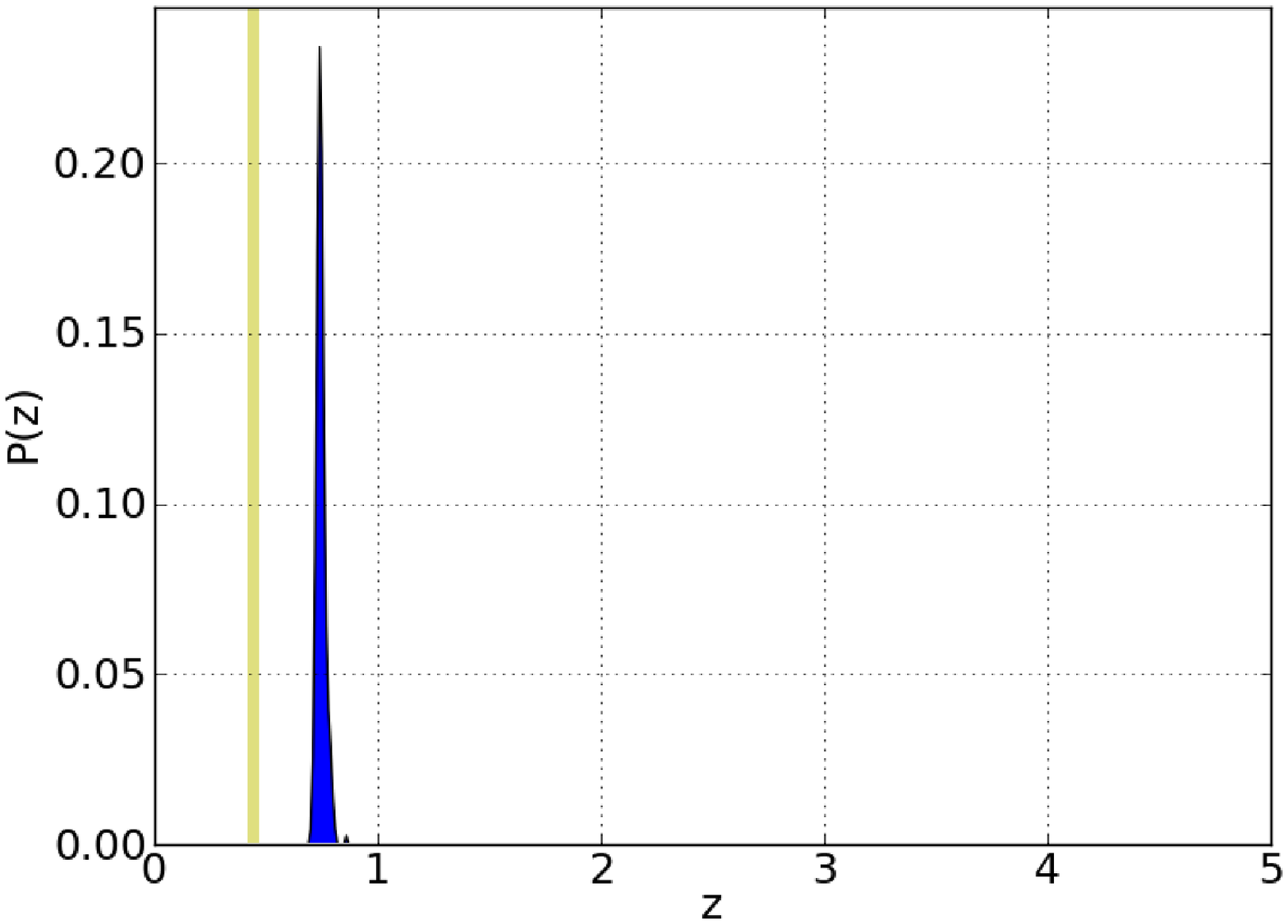}
\includegraphics[scale=0.3]{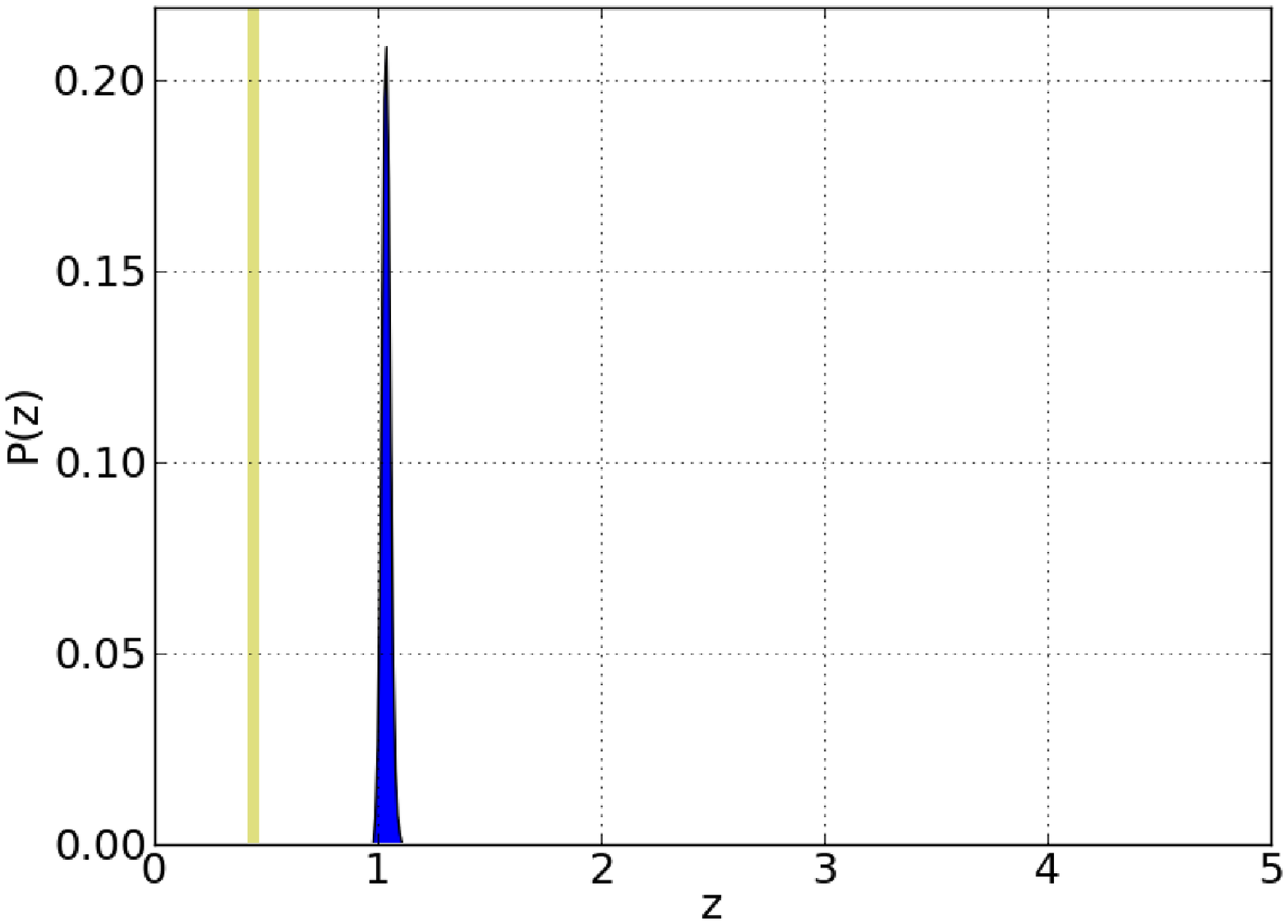}
\includegraphics[scale=0.3]{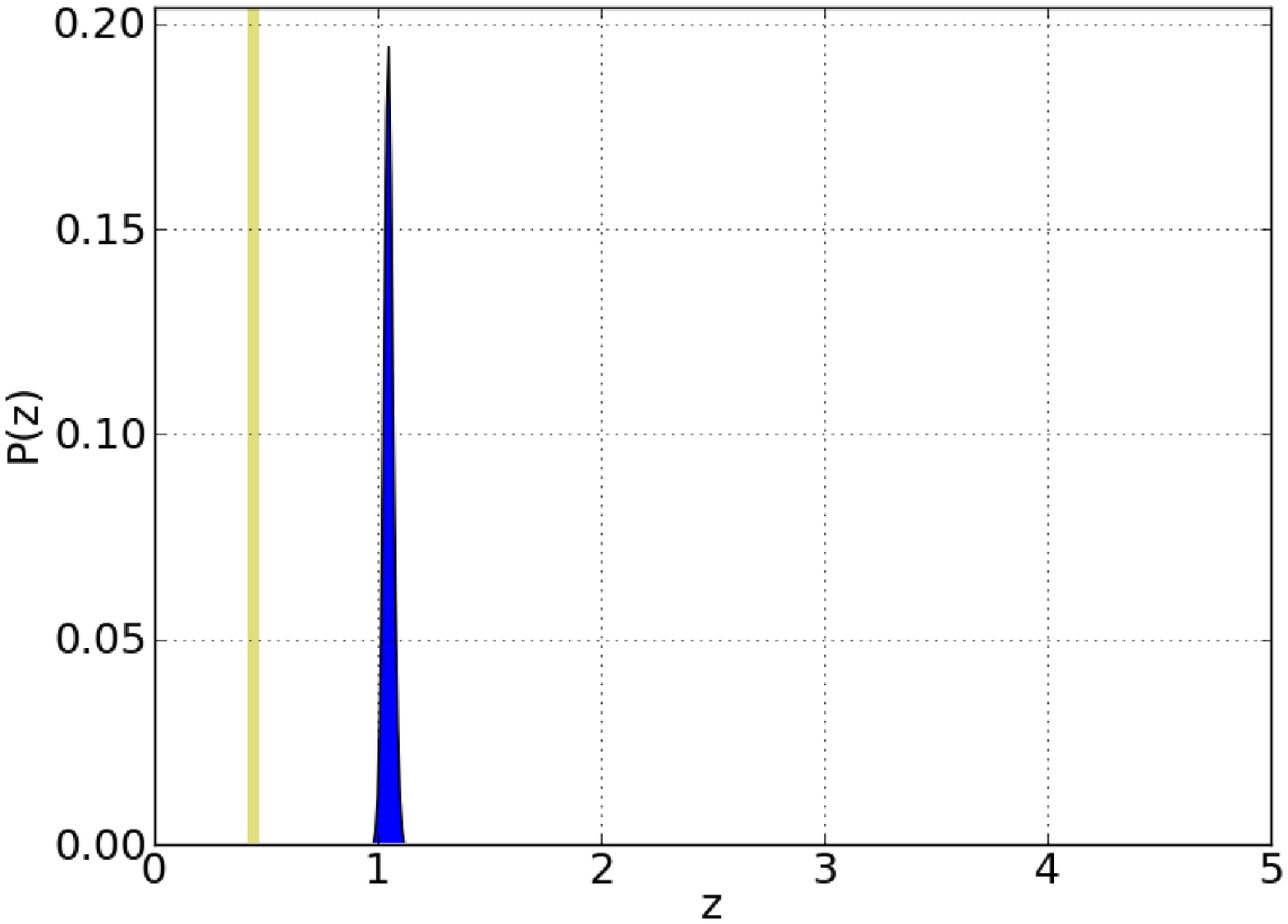}}
\hbox{
\includegraphics[scale=0.3]{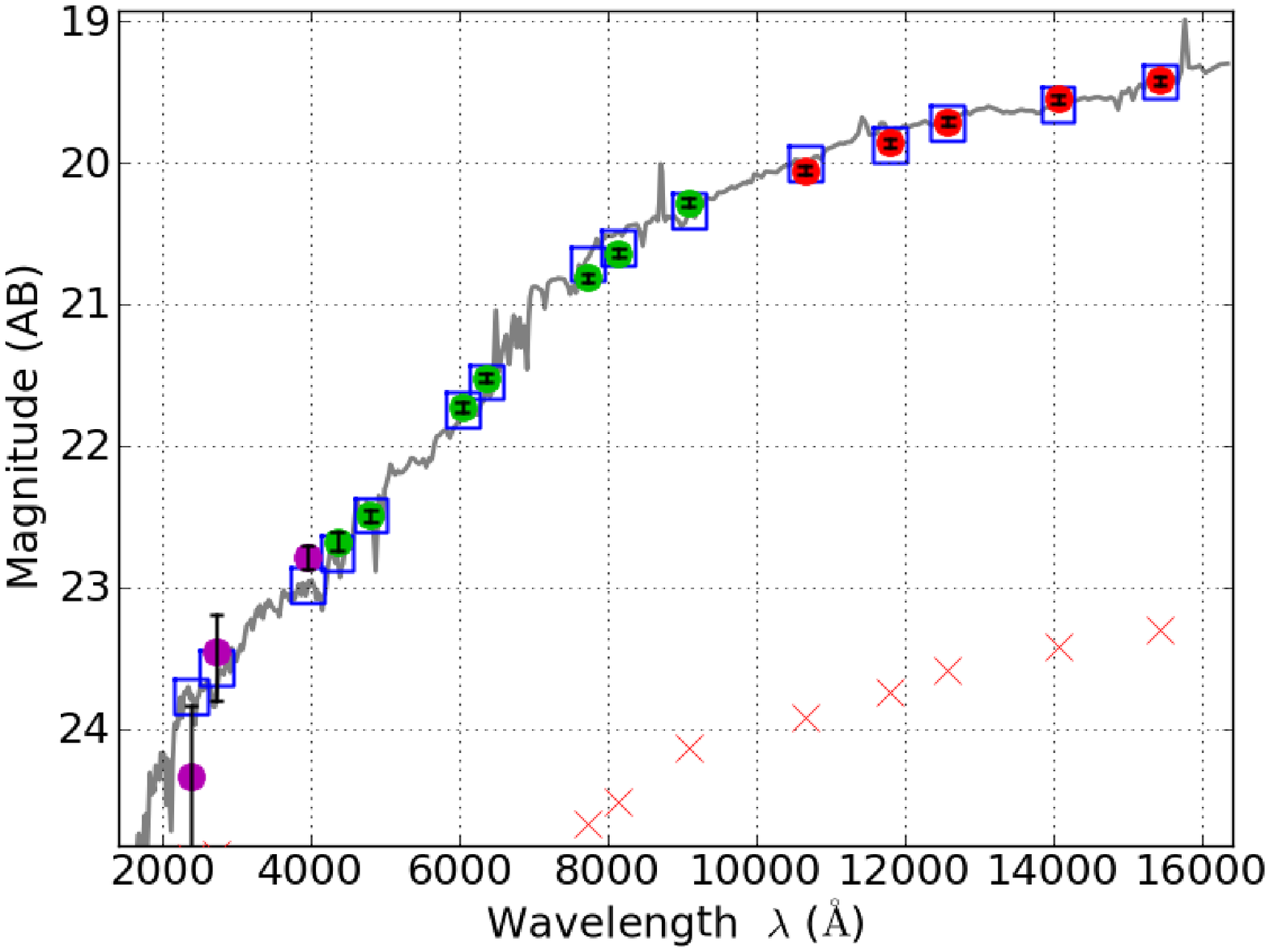}
\includegraphics[scale=0.3]{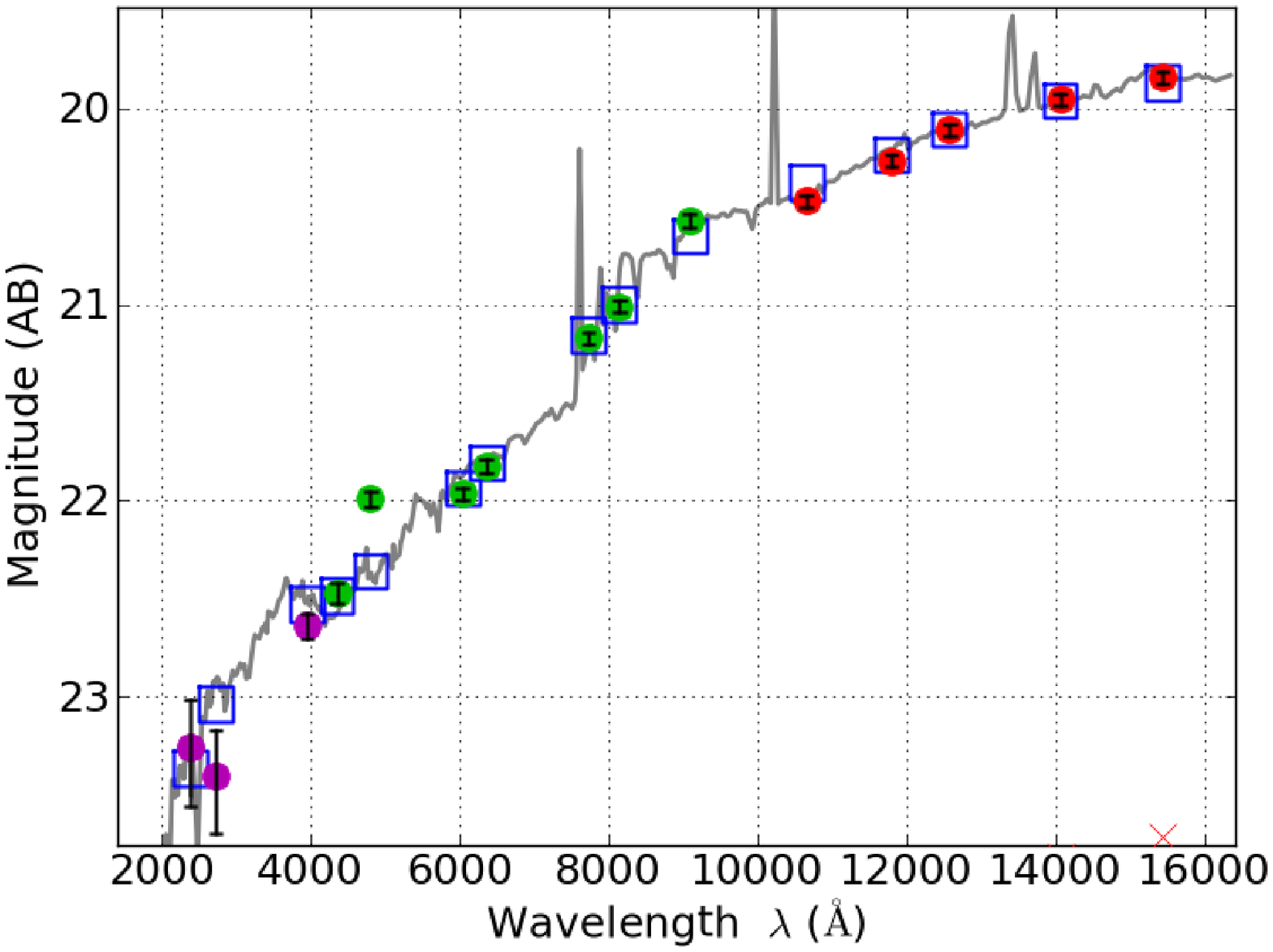}
\includegraphics[scale=0.3]{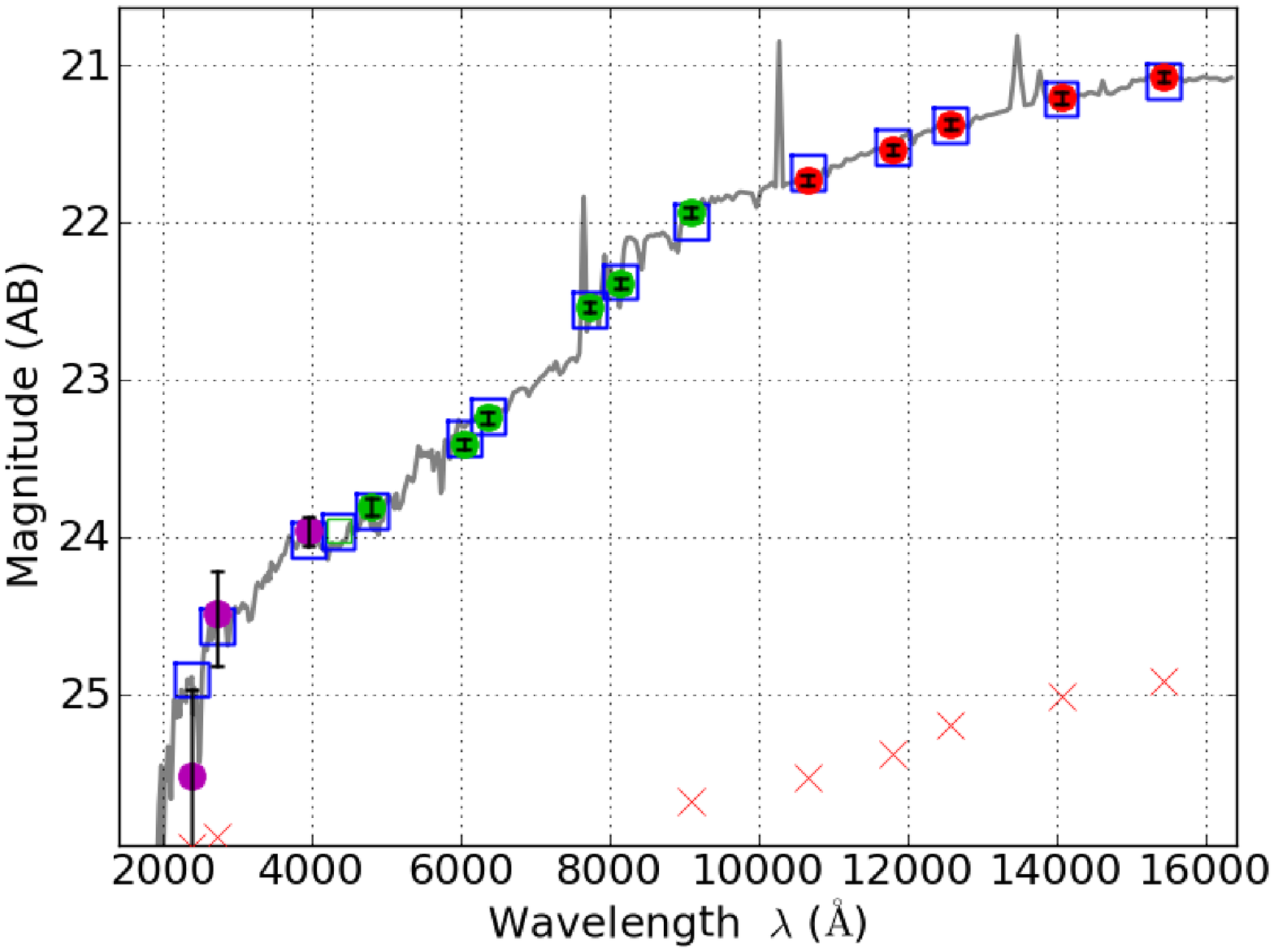}}
\caption{BPZ photometric redshift results for the strongly lensed system 1 standard photometry from 
SExtractor isophotal apertures. From left to right, figures are cut-out images of the arcs, segmentation map 
of the isophotal photometry, probability of the photometric redshift and best-fit template for the arcs considered.
From left to right, figures show arc 1.1, 1.2, and 1.3.}
\label{fig:segm_sex_sys1}
\end{figure*}

\end{document}